\documentclass[%
reprint,
superscriptaddress,
amsmath,amssymb,
aps,
prl,
]{revtex4-2}

\usepackage{amsmath}
\usepackage{amsfonts}
\usepackage{graphicx}
\usepackage{xcolor}
\usepackage{braket}
\usepackage[normalem]{ulem}

\usepackage{dcolumn}
\usepackage{bm}
\usepackage{dsfont}
\usepackage{hyperref}
\usepackage{nameref}
\usepackage{tabularx}
\usepackage{colortbl}
\usepackage{pgfmath}
\colorlet{lightgray}{white!90!black}
\definecolor{darkblue}{RGB}{0, 0, 75}
\definecolor{red}{RGB}{214, 39, 40}

\newcommand{\gradientcell}[3]{%
  \pgfmathsetmacro{\scalednum}{100 * ((#1 - #2) / (#3 - #2))}%
  \edef\tempcolor{red!\scalednum!darkblue}%
  \color{\tempcolor}\textbf{#1}
}

\newcolumntype{C}[1]{>{\centering\arraybackslash\hspace{0pt}}p{#1}}
\newcolumntype{Y}{>{\raggedright\arraybackslash}X}

\newcommand{\ML}{ML}
\newcommand{\MLIP}{MLIP}
\newcommand{\NN}{NN}

\newcommand{\PES}{PES}
\newcommand{\GA}{GA}
\newcommand{\DFT}{DFT}

\begin{document}


\title{Machine Learning and Data-Driven Methods in Computational Surface \\ and Interface Science}

\author{Lukas H\"{o}rmann}
\email{lukas.hoermann@warwick.ac.uk}
\affiliation{Department of Chemistry, University of Warwick, Gibbet Hill Road, Coventry CV4 7AL, United Kingdom}
\affiliation{Department of Physics, University of Warwick, Gibbet Hill Road, Coventry CV4 7AL, United Kingdom}
\author{Wojciech G. Stark}%
\affiliation{Department of Chemistry, University of Warwick, Gibbet Hill Road, Coventry CV4 7AL, United Kingdom}
\author{Reinhard J. Maurer}
\email{r.maurer@warwick.ac.uk}
\affiliation{Department of Chemistry, University of Warwick, Gibbet Hill Road, Coventry CV4 7AL, United Kingdom}
\affiliation{Department of Physics, University of Warwick, Gibbet Hill Road, Coventry CV4 7AL, United Kingdom}

\date{\today}

\begin{abstract}
Nanoscale design of surfaces and interfaces is essential for modern technologies like organic LEDs, batteries, fuel cells, superlubricating surfaces, and heterogeneous catalysis. However, these systems often exhibit complex surface reconstructions and polymorphism, with properties influenced by kinetic processes and dynamic behavior. A lack of accurate and scalable simulation tools has limited computational modeling of surfaces and interfaces. Recently, machine learning and data-driven methods have expanded the capabilities of theoretical modeling, enabling, for example, the routine use of machine-learned interatomic potentials to predict energies and forces across numerous structures. Despite these advances, significant challenges remain, including the scarcity of large, consistent datasets and the need for computational and data-efficient machine learning methods. Additionally, a major challenge lies in the lack of accurate reference data and electronic structure methods for interfaces. Density Functional Theory, while effective for bulk materials, is less reliable for surfaces, and too few accurate experimental studies on interface structure and stability exist. Here, we will sketch the current state of data-driven methods and machine learning in computational surface science and provide a perspective on how these methods will shape the field in the future.
\end{abstract}

\maketitle

\section{\label{sec:intro}Introduction}

Many modern technological challenges crucially depend on the properties of surfaces and interfaces. This includes the control of charge and energy transfer across electrode/electrolyte interfaces in batteries~\cite{leung2020dft} and fuel cells,~\cite{stamenkovic2017energy} the characterization of structure and dynamics of wear and lubrication at tribological interfaces,~\cite{baykara2018emerging} the optimization of chemical transformations at metal surfaces in heterogeneous catalysis,~\cite{somorjai1994surface} corrosion science~\cite{maurice2018progress} and surface functionalization.~\cite{wieszczycka2021surface} Surface/thin film growth processes such as atomic layer deposition, chemical vapor deposition, or molecular beam epitaxy are highly industrially relevant. Increasing demand for functional thin-films and surface nanostructures also increases their structural complexity. As most modern materials involve multiple components, understanding the structure and properties of thin films, composites, buried interfaces, and exposed surfaces is now more important than ever. The atomic-scale characterization and design of functional interfaces require understanding and manipulation at the nanoscale, which often cannot be delivered by experimentation alone.

Computational simulation of surface and interface processes has become central to modern surface science. Few fields rely as strongly on the synergy between atomistic simulation and experimental study. This synergy is achieved by minimizing the gap between experimental complexity and simulation models, often through the use of model surfaces and ultra-high vacuum conditions, which, for example, enable atomic resolution to be achieved in scanning probe microscopy. However, surfaces in real-world applications are often more complex, featuring defects and partial disorder. Additionally, ambient pressure and interacting molecules play crucial roles in many applications, such as catalysis. By advancing the study of complex surface systems and dynamic processes at large length- and time scales with high throughput, machine learning (\ML) and data-driven approaches have the potential to bring atomistic simulation and experiment even closer, offering improved mechanistic understanding of surface dynamics, reaction pathways, growth processes, and mechanical and electronic properties.

\begin{figure*}
	\centering
	\includegraphics[width=1\linewidth]{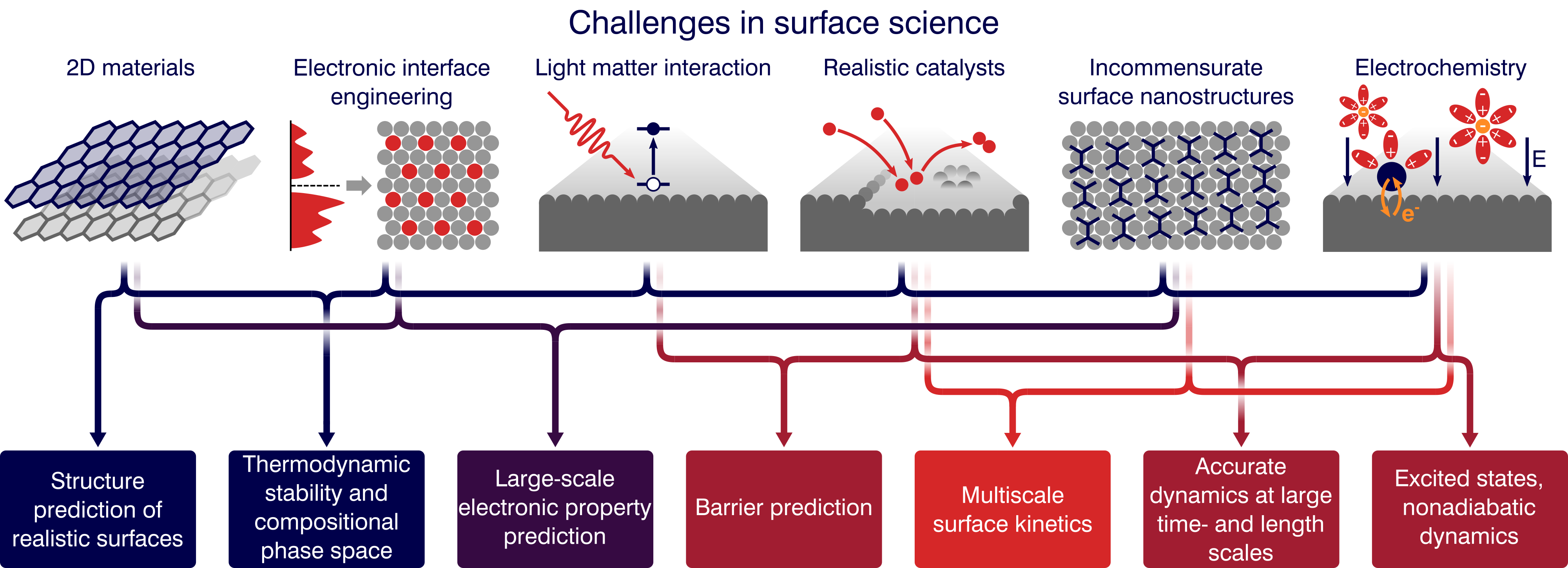}
	\caption{Schematic depiction of exemplary grand challenge areas in surface science from left to right: two-dimensional materials, description of realistic catalysts, incommensurate surface nanostructures, light-matter interaction at interfaces, electronic interface engineering, surface electrochemistry; Frontier modelling requirements arise from these application challenges, which are listed in the bottom panel.}
	\label{fig:surfacesciencechallenges}
\end{figure*}

Common {\ML} methods used in surface science encompass neural networks ({\NN}), Bayesian regression methods, decision trees, support vector machines, and genetic algorithms. Citations for specific uses are given in later sections. These methods can learn expressions for formation energies, potential energy surfaces (\PES), and other properties, provide frameworks to efficiently explore the configuration space of the material, or facilitate the optimization of a target property. {\MLIP}s, in particular, are highly impactful and have revolutionized the simulation of bulk materials and are in the process of taking over chemical and biomolecular simulations.~\cite{deringer2019machine, mishin2021machine} {\MLIP}s are omnipresent in surface science and will be discussed throughout this review. Considering the broad range of {\ML} applications in surface science, we do not dedicate a separate chapter to {\MLIP}s. Instead, we discuss them as part of the section on \hyperref[sec:dynamics_at_surfaces]{Accurate dynamics at large time- and length scales}, as MLIPs are most commonly used to accelerate dynamics simulations.

\paragraph*{\textbf{Motivation of the review.}}

A number of comprehensive reviews have been published on {\ML} from different perspectives such as heterogeneous catalysis~\cite{kitchin2018machine, chen2020computational, toyao2019machine, yang2019machine} and experimental surface science~\cite{kalinin2021automated, gordon2020machine}. This review targets computational surface scientists seeking to integrate {\ML} techniques, providing an overview of current capabilities and limitations. Surfaces and interfaces present a unique challenge due to complex processes such as charge transfer and bond formation as well as competing interactions such as covalent, electrostatic, dispersion forces, and the coexistence of localized and delocalized electronic states in large unit cells with hundreds or thousands of atoms.~\cite{hofmann2021first} Examples such as the CO on metals puzzle~\cite{gajdovs2005co, feibelman2001co} demonstrate that semi-local Density Functional Theory (DFT) often falls short in accurately describing surfaces and interfaces.

Current challenges in surface science (sketched in Figure \ref{fig:surfacesciencechallenges}), such as the modeling of two-dimensional materials, electronic interface engineering, light-matter interaction at interfaces, description of realistic catalysts, incommensurate surface nanostructures, surface electrochemistry, provide specific demands on computational modelling capabilities, which machine learning and data-driven methods can help to address.

In this review, we focus on these broader computational simulation challenges spanning gas-surface, solid-liquid, and solid-solid interfaces. While each system presents unique challenges, many computational workflows are applicable across all types of surfaces and interfaces. We explore how new {\ML}-enabled workflows are transforming these surface science applications, focusing on their role in (A) structure prediction of realistic surfaces, (B) thermodynamic stability and compositional phase space under realistic conditions, (C) large-scale electronic property prediction, (D) barrier prediction, (E) multiscale surface kinetics, (F) accurate dynamics at large time- and length scales, and (G) excited states, surface spectroscopy and nonadiabatic dynamics. We address the unique challenges of studying lower-dimensional systems compared to bulk or molecular systems and identify gaps in {\ML} tools that, if filled, could greatly enhance computational surface science.

\section{Structure prediction of realistic surfaces}
\label{sec:surface_structure_and_stability}

Understanding the structure of an interface, surface or adsorbed layer is crucial for studying properties such as charge transfer, surface states, level alignment, or interface dipole. Surface structure prediction is challenging due to the vast number of possible structures and the computational expense of large unit cells containing hundreds of atoms. Advanced optimization methods and {\ML} approaches have enabled effective local and global structure optimization and efficient sampling of the {\PES}.

\paragraph*{\textbf{Local structure optimization.}}

Today, local structure optimization in surface science is considered a mostly solved issue. Common local optimization routines such as the gradient descent, Broyden–Fletcher–Goldfarb–Shanno, or Nelder-Mead methods can often be used in conjunction with first-principles computations. For particularly demanding problems, surrogate models can be employed: For example, using a Bayesian {\ML} approach, Garijo del R{\'\i}o et al.~\cite{del2019local, garijo2020machine} accelerated local geometry optimizations of CO on Ag(111) and C on Cu(100). Their algorithm, called \href{https://gitlab.com/egarijo/bondmin}{GPMin}, conducts optimizations on a machine-learned {\PES} and improves the underlying {\ML}-model by adding each newly found local minimum to the training data. GPMin is available in the atomic simulation environment (ASE).~\cite{larsen2017atomic} 

\paragraph*{\textbf{Global structure optimization.}}

Unlike local optimization, global optimization is significantly more challenging due to the vastly larger search space. The uniquely complex interactions at surfaces and interfaces (charge transfer, hybridization, level alignment, etc.) render purely first-principles approaches intractable.~\cite{hofmann2021first, hormann2019sample} {\ML} based approaches, such as Bayesian regression, tree-based algorithms, genetic algorithms ({\GA}s), and {\NN}s have enabled significant advances in tackling complex systems with many degrees of freedom.

For instance, the group of Bj{\o}rk Hammer developed \href{https://gofee.au.dk/.}{GOFEE} (global optimization with first-principles energy expressions),~\cite{bisbo2020efficient} which employs a Gaussian process regression (GPR) based surrogate model for energies. GOFEE makes efficient use of limited first principles data through adaptive sampling (see \hyperref[sec:dynamics_at_surfaces]{Accurate dynamics at large time- and length scales}) and was applied to study the oxidation and oxygen intercalation of graphene on Ir(111). Similarly, ~Kaappa et al. used Gaussian processes to create a surrogate model for the {\PES} to enable global optimization. Their algorithms, BEACON (Bayesian Exploration of Atomic Configurations for Optimization)~\cite{kaappa2021global} and ICE-BEACON~\cite{kaappa2021atomic}, are available in the \href{https://gitlab.com/gpatom/ase-gpatom/}{GPAtom} package. Another approach successfully used for global optimization is Gaussian approximation potential (GAP)~\cite{bartok2010gaussian}. For example, Timmermann et al. applied GAP with simulated annealing to optimize low-index surfaces of rutile (IrO\textsubscript{2}).~\cite{timmermann2020iro}

A different approach was used by Li et al.~\cite{li2022transferable} to create a transferable {\ML} model for the prediction of adsorption energies of single-atom alloys. They utilized \href{https://github.com/dmlc/xgboost}{XGBoost},~\cite{chen2016xgboost} which uses a boosted tree algorithm based on gradient boosting. Hereby a series of trees is used: The first tree learns the original data, while every subsequent tree learns the residual of the previous tree, thereby improving the accuracy. XGBoost is widely used in computational materials research and has, for instance, been employed to model molecular adsorption in metal-organic framework~\cite{abdi2021modeling} or for the discovery of heterogeneous catalysts.~\cite{suzuki2019statistical}

{\GA}s have been widely applied for global optimization of surface structures. While many of these algorithms were originally developed to find minimum energy structures for bulk materials or proteins, they have since been adapted for surfaces. A prime example is \href{https://uspex-team.org/en}{USPEX} (Universal Structure Predictor Evolutionary Xtallography)~\cite{glass2006uspex} which uses a {\GA} to computationally predict crystal structures. Wen et al. used the USPEX code to determine the structure of a mixed-metal oxide monolayer grown at an oxide support.~\cite{wen2016fe} A key challenge in performing {\GA} optimizations with first-principles calculations is the high computational cost of surface calculations, prompting efforts to minimize the required computations. Chuang et al. developed a {\GA} that avoids the evaluation of duplicate structures by ensuring that the structures in the pool either differ in energy or atomic displacements, allowing them to determine the reconstructions of semiconductor surfaces.~\cite{chuang2004finding} Computational effort can also be reduced by coupling first-principles calculations with less resource-intensive methods. For example, Bjørk Hammer's group implemented a two-stage {\DFT} optimization approach, which pre-screens candidate structures with less accurate, lower-cost methods to identify and remove duplicates. Promising new structures are then refined using more accurate DFT calculations.~\cite{vilhelmsen2012systematic, vilhelmsen2014genetic} Another efficient strategy is to pair a {\GA} with a surrogate model for the target property. Jacobsen et al. used a kernel ridge regression {\ML} model with adaptive sampling to guide a {\GA} for optimizing surface reconstructions of oxide materials,~\cite{jacobsen2018fly} leveraging feature vectors from Oganov et al.~\cite{oganov2009quantify} with adaptive sampling (see \hyperref[sec:dynamics_at_surfaces]{Accurate dynamics at large time- and length scales}) to generate training data.

Particle swarm optimization has also been applied to predict surface structures. A notable example is the CALYPSO (Crystal structure AnaLYsis by Particle Swarm Optimization) code~\cite{wang2012calypso}. Using CALYPSO, Lu et al. developed a method to explore the surface structures such as diamond surface reconstructions featuring self-assembled carbon nanotubes arrays~\cite{lu2014self}. The code has also been employed to predict solid-solid interface structures. Optimal lattice-matched superlattices are determined, employing constraints on interatomic distances and atomic coordination numbers to generate starting interface structures~\cite{gao2019interface}. Moreover, CALYPSO was used to reveal the CeO\textsubscript{2} surface reconstruction. First-generation structures were created with CALYPSO, based on input parameters such as the bulk crystal surface structure, number of atoms that form the interface, and thickness of reconstructed layers~\cite{zhang2024ceo2}. {\DFT} was used to optimize these structures compared to experiment.

While most {\ML} applications in surface science use supervised learning methods, reinforcement learning methods are also gaining traction in the structure prediction community. Meldgaard et al. predicted crystal surface reconstructions by using reinforcement learning combined with image recognition.~\cite{meldgaard2020structure} The learning agent employs a deep {\NN} to decide the placement and type of the next atom in an incomplete structure. The new structure -- which represents a new state -- is evaluated using {\DFT} and a reward is determined. States and rewards are saved and used by the agent for future atom placements. This algorithm is an extension (to 3D structures) of the ASLA method~\cite{jorgensen2019atomistic} which generated 2D materials and molecules with reinforcement learning.

\paragraph*{\textbf{Open challenges.}}

Despite major advances in accurately predicting global minimum surface structures, current methods are limited to systems with a few hundred atoms.~\cite{hu2021forcenet, todorovic2019bayesian} Large-scale systems, such as incommensurate surface structures or realistic catalytic surfaces, containing tens of thousands of atoms remain beyond reach due to the limited efficiency and precision of current inference methods. 

Surfaces regularly exhibit multiple (even thousands) polymorphs with differences in stability below $1~kcal/mol$ per atom ($40~meV$),~\cite{hormann2019sample, hormann2022bistable} despite exhibiting different properties, such as the work function. DFT is the workhorse method in the field, but different functionals, especially the typically used Generalized Gradient Approximation (GGA), struggle to yield energy computations that are consistently within an accuracy of $1~kcal/mol$ per atom.~\cite{hormann2020reproducibility, maurer2016adsorption, carbogno2022numerical} While more accurate first-principles methods exist, their cost conflicts with the demand for large datasets to improve {\ML} model accuracy. More data-efficient approaches are required. Several examples of transfer learning exist for molecules~\cite{zaverkin2023transfer, smith2018less} and bulk materials.~\cite{rowe2020accurate, deringer2020general} Recently, foundation models (also called universal {\MLIP}s) have been introduced, that are trained on large and diverse databases and often deliver adequate accuracy for tasks such as predicting formation energies and performing preliminary geometry optimizations. 

Nevertheless, gaps in accuracy and amount of available training data as well as limited inference efficiency remain open challenges in the prediction of realistic surface structures.

\begin{figure}
	\centering
	\includegraphics[width=1\linewidth]{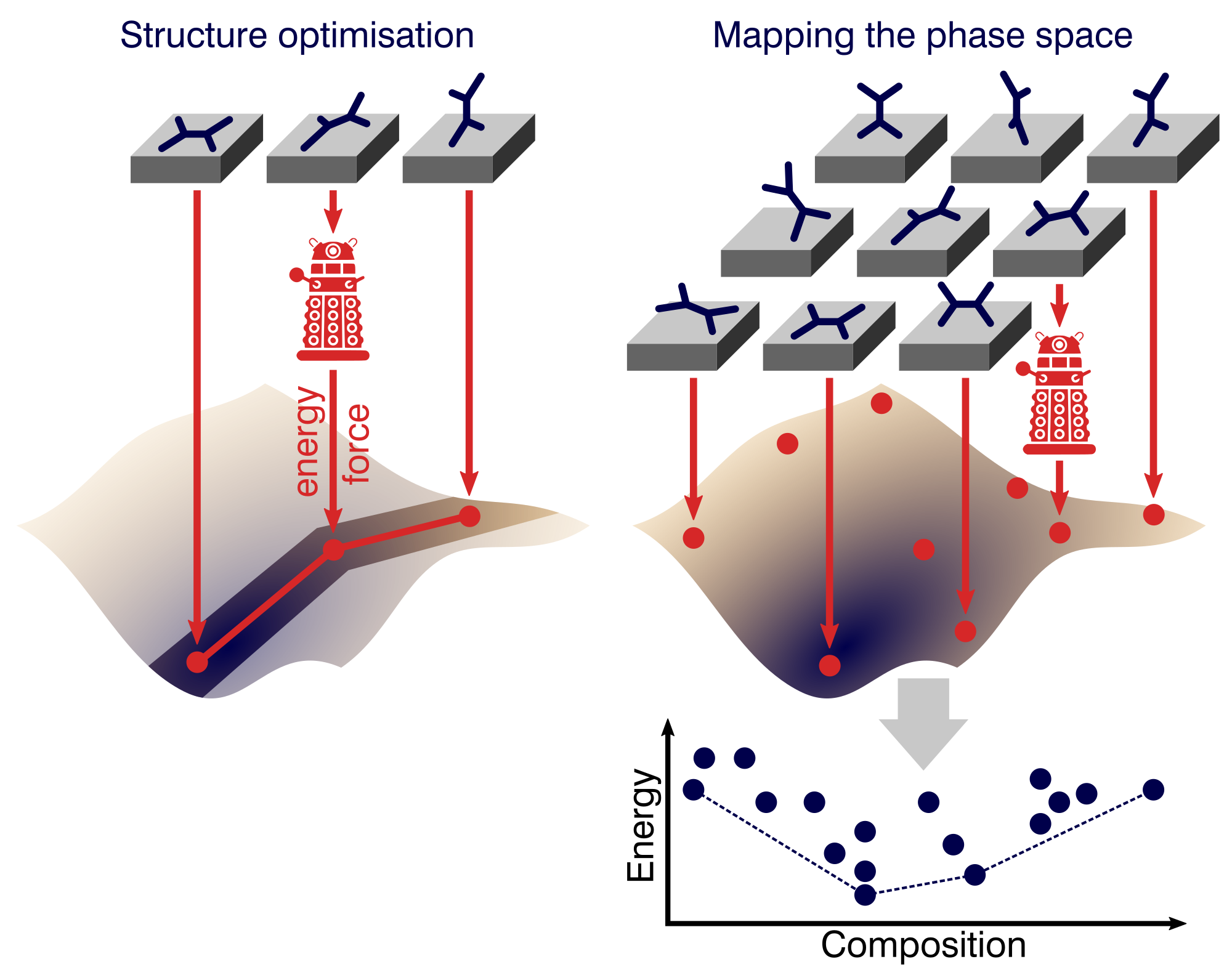}
	\caption{Schematic representation of structure search tasks: (A) For local and global structure optimization, only part of the {\PES} is sampled. (B) The phase space can be mapped by determining the entire {\PES}.}
	\label{fig:methodstimeline}
\end{figure}

\section{Thermodynamic stability and compositional phase space under realistic conditions}

Understanding the thermodynamic stability and compositional phase space enables the prediction of the most likely surface structures and compositions under realistic environmental conditions, such as temperature, pressure, and chemical potentials. This insight is essential for the targeted design of new materials and enhances the interpretation of experiments conducted under these conditions. Exploring thermodynamic and compositional phase spaces requires mapping the entire configuration space (i.e., the {\PES}).

The determination of {\PES}s and adsorption geometries of single molecules on surfaces was one of the first applications of {\ML} in surface science. An important contribution is the BOSS code,~\cite{todorovic2019bayesian} which combines GPR with descriptors based on Cartesian and internal coordinates, using model uncertainty for active learning (see \hyperref[sec:dynamics_at_surfaces]{Accurate dynamics at large time- and length scales}). BOSS requires only a few hundred energy evaluations to construct a five-dimensional {\PES}, though the number of necessary training data points increases exponentially with the degrees of freedom. A similar approach was developed by H{\"o}rmann et al., using radial distance functions as descriptors to predict the {\PES} of single molecules on surfaces.~\cite{hormann2020reproducibility} The authors later extended this method to continuous adsorbate layers and predicted other properties, such as work functions for near-incommensurate surface structures.~\cite{hormann2022bistable, hormann2023impact} The same authors also developed a surface structure search algorithm called SAMPLE.~\cite{PhysRevMaterials.2.043803, hormann2019sample} SAMPLE can learn energies and work functions~\cite{jeindl2022much} and generate a comprehensive list of surface structures based on coverage and the number of molecules per unit cell. It used a pair-wise potential fitted using Bayesian linear regression from a training set of a few hundred DFT evaluations. The training set is chosen using optimal design theory to maximize the information gained, addressing a key challenge in {\ML} for surface science—efficient data use. Combined with ab-initio thermodynamics~\cite{rogal2007ab}, SAMPLE can provide access to thermodynamic quantities via the partition function. A similar approach to SAMPLE is the GAMMA (Generalized block AsseMbly Machine learning equivalence clAss sampling modeling) approach~\cite{packwood2017chemical}. It employs an Ising-type model based on energy calculations of molecule-molecule and molecule-substrate interactions performed in isolation. This approach contrasts with SAMPLE, which derives these interactions from formation energy calculations of the adsorbed system. GAMMA explores the configuration space stochastically using an equivalence class sampling algorithm that removes redundant information, allowing it to consider large unit cells. 

A different approach was taken by Ulissi et al. who used GPR to learn the coverage-dependent free energy for IrO\textsubscript{2} and MoS\textsubscript{2} surface reconstructions.~\cite{ulissi2016automated} Training on a small number of {\DFT} calculations, selected using the inherent uncertainty of GPR, allows the authors to construct Pourbaix diagrams that illustrate the stability of surfaces in electrochemistry applications as a function of pH and electrochemical potential.

{\ML} has also facilitated research into the phase behavior of solid-solid interfaces. Moayedpour et al., for instance, studied epitaxial interfaces of tetracyanoquinodimethane on tetrathiafulvalene~\cite{moayedpour2023structure}. The authors used an improved version of the Ogre~\cite{yang2020ogre} code to generate possible surface structures and predicted energies using the ANI~\cite{smith2017ani} {\MLIP} in conjunction with the D3~\cite{grimme2011effect} vdW-correction. The Ogre code facilitates generating ideal (perfectly ordered) interfaces, without considering growth conditions. Huber et al. have used gradient-boosted decision trees, based on descriptors which depend only on local properties of the grain boundary, to predict the segregation energy distribution, and thus the segregation isotherm of grain boundaries of metallic solutes in aluminium.\cite{huber2018machine} Their model provides improved predictive power over the common Langmuir–McLean model.

A data-driven, but not directly {\ML} related approach for determining surface thermodynamic properties is nested sampling~\cite{skilling2004nested}. Yang et al. used this method to calculate adsorbate phase diagrams, incorporating all relevant configurational contributions to the free energy.\cite{yang2024surface} Nested sampling requires significant numbers of energy evaluations, the cost of which was overcome by the authors by using a Lennard-Jones potential to describe interactions.

Ulissi et al. used GPR to learn the coverage-dependent free energy for IrO\textsubscript{2} and MoS\textsubscript{2} surface reconstructions.~\cite{ulissi2016automated} Training on a small number of {\DFT} calculations, selected using the inherent uncertainty of GPR, allows the authors to construct Pourbaix diagrams that illustrate the stability of surfaces in electrochemistry applications as a function of pH and electrochemical potential.

\paragraph*{\textbf{Open challenges.}}

Exploring thermodynamic stability and compositional phase space presents similar challenges to predicting surface structures (see \hyperref[sec:surface_structure_and_stability]{Structure prediction of realistic surfaces}). Studying thermodynamic stability presents the additional challenge of learning properties such as the vibrational enthalpies and free energies, which require accurate modeling of anharmonic effects, electronic effects, and environmental conditions. Most {\ML} applications in surface thermodynamics focus on predicting energies of phase candidates. While progress has been made in learning thermodynamic properties for molecular or bulk systems~\cite{riniker2017molecular, manzoor2020predicting}, as well as {\ML}-enhanced CALPHAD (Calculation of Phase Diagrams) modeling for bulk materials,\cite{zeng2021revealing, liu2022design} little work exists on directly learning surface vibrational enthalpies or other thermodynamic properties.

Comprehensive, accurate databases for vibrational eigenvalues and eigenvectors, phase transitions, or other thermodynamic properties of surface structures are lacking. While thermodynamic properties such as the adsorption energies or vibrational properties can be found in common databases (see Table \ref{tab:databases}) the amount of data that exists on  surfaces is limited. For instance, the recently published Molecular Vibration Explorer~\cite{koczor2022molecular} in the materialscloud~\cite{talirz2020materials} is only dedicated to a specific class of molecule. As a result of this data gap, little work exists on benchmarking {\ML} approaches for thermodynamic properties.

\section{Large-scale electronic property and spectroscopy prediction}
\label{sec:electronic_property}

The prediction of electronic properties for large-scale systems is a crucial aspect of characterizing and tailoring the properties of realistic surfaces and interfaces with defects, step edges and grain boundaries -- all of which contribute to measurable electronic transport, reactivity, and spectroscopic signatures. ML surrogate models of electronic structure and spectroscopic properties are much less developed than {\ML} models for energy and force prediction. To date only few proof-of-principle models have been applied in the context of surface science. Their obvious applications lie in high throughput property prediction and in overcoming the intrinsic scaling limitations of DFT.

\paragraph*{\textbf{Learning level alignment, molecule-surface hybridisation, band bending.}}
Scalar electronic properties can often be learned with methods originally designed to learn energies. For instance, the SAMPLE code~\cite{hormann2019sample}, is also able of predicting the work function. Choudhary et al. followed a similar philosophy when developing InterMat~\cite{choudhary2024intermat}. This approach uses {\DFT} and {\NN}s to predict band offsets of semiconductor interfaces. A graph {\NN} model is used to predict the valence band maxima and conduction band minima of the respective surfaces, using the atomic structures as input data. Training data was generated with {\DFT}. By aligning the vacuum levels of these surfaces, the band offset is determined (according to Anderson's rule).

Gerber et al. used a purely data-driven approach for InterMatch~\cite{gerber2023high}, a high-throughput computational framework designed to efficiently predict charge transfer, strain, and superlattice structures at material interfaces. The algorithm leverages databases of individual bulk materials to streamline the prediction process. By accessing lattice vectors, density of states, and stiffness tensors from the Materials Project, InterMatch estimates interfacial properties. The code was used to study transition metal dichalcogenides and qualitatively reproduced {\DFT} simulations and experiment. This approach enables high throughput pre-screening.

\paragraph*{\textbf{Hamiltonian and band structure prediction.}}

Nonlinear learning techniques to parameterize Hamiltonians have a long legacy in semi-empirical and tight-binding models and several end-to-end learning strategies of Slater-Koster integral tables from calculated properties have been proposed for molecules and materials.~\cite{Li2018, schattauer_machine_2022, jorner_ultrafast_2024} Beyond the limitations of the 2-center integral approximations, various linear, kernel-based,~\cite{Hegde2017} and deep-learning-based representations of electronic structure have been proposed for molecules.~\cite{schutt2019unifying, gastegger2020deep} These models typically target the Kohn-Sham Hamiltonian in local orbital representation, which provides an atom-centered representation that can encode the rotational transformation properties of local orbital integrals~\cite{unke_se3-equivariant_2021} and does not suffer from the non-smoothness and phase issues associated with learning eigenvalues and eigenvectors directly~\cite{westermayr_machine_2021}. Additionally, this representation can also be extended to condensed phase systems as the Bloch transform of the real-space Hamiltonian readily enables the calculation of the band structure and electronic density of state (DOS). Examples of recent models include the linear ACEhamiltonians model based on the Atomic Cluster Expansion \cite{zhang_equivariant_2022}, DeepH,~\cite{li_deep-learning_2022} and DeepH-E3 models~\cite{gong_general_2023}. The latter has been applied to two-dimensional materials such as twisted graphene bilayers. 

\paragraph*{\textbf{Prediction of electronic density of states, the electron density, and spectroscopic properties.}}

The ability to predict the electronic DOS or the electron density directly (rather than through diagonalisation of a Hamiltonian) is valuable in studying charge transfer and the relationship between surface structure and electronic properties.
SALTED is an approach to learning the electron density expressed as a linear combination of atom-centred basis functions. The corresponding basis coefficients are trained using Gaussian process regression.~\cite{lewis_learning_2021} This approach has been applied to condensed phase systems, two-dimensional materials and electrochemical interfaces. \cite{Grisafi_2024} In the latter case, it is able to accurately predict the charge distribution of a metal electrode in contact with an explicit aqueous electrolyte.
Several approaches have been proposed that represent the DOS in condensed matter systems either by directly learning a spectrum or by representation in a suitable atom-centred basis.~\cite{Chandrasekaran2019, ben_mahmoud_learning_2020, kong_density_2022} This approach has been successfully employed for Al metal slabs and is readily applicable in surface science applications.

Related to learning the DOS and electron density, are {\ML} models for spectroscopic properties. Methods such as ultraviolet photoelectron spectroscopy (UPS), x-ray photoelectron spectroscopy (XPS), and near edge X-ray absorption fine structure  (NEXAFS) provide invaluable information on the chemical state and bonding environment of atoms in samples. Often spectroscopic assignment is challenging and first principles simulations can provide valuable input. {\ML} methods provide the means to accelerate predictions, with kernel ridge regression and neural network models having been proposed to predict (inverse) photoemission signatures,~\cite{westermayr_machine_2021} XPS,~\cite{aarva_understanding_2019, golze_accurate_2022} and NEXAFS~\cite{rankine_deep_2020, kotobi_integrating_2023}, but have mostly focused on bulk system.  Additionally, {\ML} approaches enable novel discoveries, such as inverse structure determination based on target spectra~\cite{zarrouk_experiment-driven_2024}. ML methods have seen widespread application in spectroscopy, which has been summarized in comprehensive reviews on the subject.~\cite{westermayr_chemrev_2021, rankine_progress_2021, meza2021applications, zhang2022brief}

\paragraph*{\textbf{Open Challenges.}} A critical open challenge in electronic structure and spectroscopy surrogate models and their applications to surface science is the integration with existing tools and workflows. Most current models are focused on bulk systems, but would easily be transferable to surface science. However, there exists a lack of sufficient training data. So far, little work has gone into making proof-of-principle models user-friendly, scalable and well-integrated with electronic structure and simulation software. General future directions have been previously articulated~\cite{westermayr_perspective_2021} and integration with dynamics will be covered in \hyperref[sec:nonadiabatic]{Excited states and nonadiabatic dynamics}. Specifically in the context of surface science, these models will face similar challenges as first principles and semi-empirical methods as interfaces offer a rich diversity in local atom and bond environments that is hard to capture.~\cite{hofmann2021first} In addition, long-range electrostatic and dispersion interaction contributions to the electronic properties must be considered. We believe that the developments in this field will drastically pick up in the coming years with many {\ML} developments, such as equivariant NNs, straightforwardly transferring into electronic structure surrogates.

\section{Reaction barrier prediction}
\label{sec:kinetics}

Most experiments and industrial applications (e.g., surface deposition processes, catalysis) operate away from thermodynamic equilibrium, making it essential to consider kinetic processes at surfaces. This involves identifying metastable states, barriers, transition states, and transition rates at surfaces. The extensive energy and force evaluations needed for such studies demand highly efficient {\ML} approaches.

\paragraph*{\textbf{Transition state search}}

Transition state search is essential to the study of surface dynamics. The transition energy, for instance, defines reaction rates. Although many successful algorithms were developed for transition path and transition state search, like e.g. nudged elastic bands (NEB), there is still an urgent need for automation and speed-up. Multiple novel methods~\cite{behler2017first, zhang_embedded_2019, batzner20223, schutt2017schnet, schutt2018schnet, bartok2010gaussian, drautz_atomic_2019, khorshidi_amp_2016, vandermause_--fly_2020} allow constructing {\MLIP}s based on first principles data. Such {\MLIP}s can replace more computationally expensive DFT calculations within the NEB algorithm. In most cases, training {\MLIP}s during the evaluation of NEBs is as computationally costly as directly evaluating the NEBs using DFT codes. However, the resulting models can be later used for other purposes, e.g. for kinetics. One of the earliest examples of {\ML} accelerated transition state search was introduced by Peterson~\cite{peterson_acceleration_2016}, who created an approximate {\NN}-based {\PES} using the atomistic {\ML}-based package (Amp)~\cite{khorshidi_amp_2016} to speed up NEB calculations. After finding an initial saddle point, results are confirmed with first-principles calculations, and the model is retrained with the new data points. This is repeated until agreement with the first-principles calculations is reached. The method was employed for two systems: diffusion of a gold atom on an Al(100) surface infused with Pt atoms and for bond rotation in ethane, in both cases requiring significantly fewer force calls than standard DFT-based NEB. 
Schaaf et al. presented a general protocol for the prediction of energy barriers of catalytic systems by training {\MLIP}s using active learning based on the energy uncertainty of individual atoms. The protocol was applied to the conversion of CO\textsubscript{2} to methanol at In\textsubscript{2}O\textsubscript{3} surface with a single oxygen vacancy.~\cite{schaaf_accurate_2023} 

A recent effort toward large pre-trained {\MLIP}s capable of predicting reaction barriers across diverse catalytic systems was enabled by the OC20NEB database~\cite{wander_cattsunami_2024}, containing 932 NEB calculations at the GGA-DFT level. Wander et al. used this database within the CatTSunami framework to validate {\NN}-based {\MLIP} models trained on the OC20 database~\cite{chanussot2021open} (see \hyperref[sec:dynamics_at_surfaces]{Accurate dynamics at large time- and length scales}), based on 153M and 31M parameter Equiformer v2~\cite{liao_equiformerv2_2024}, GemNet-OC~\cite{gasteiger2021gemnet}, PaiNN~\cite{schutt_equivariant_2021}, and DimeNet++~\cite{klicpera2020fast} models.~\cite{wander_cattsunami_2024} The results of this study are reproduced in Table~\ref{tab:neb_oc20}, which shows convergence and success rate of predicting barriers of molecules detaching from the surface (desorption), dissociating at a surface (dissociation) or atom exchange between two reactants (transfer). Equiformer v2 shows the best performance, even using the lighter (31M) model. PaiNN and DimeNet++ underperformed in particular for the dissociation and transfer reactions, wherein more degrees of freedom participate in the reaction than for desorption reactions.

\begin{table}[]
    \centering
    \setlength{\tabcolsep}{0pt}
    \begin{tabular}{c|C{2.5cm}|C{2.3cm}|C{1.8cm}}
    \textbf{Method} & \textbf{Reaction type} & \textbf{\% converged} & \textbf{\% success} \\
    \hline
    Eq2 (153M)~\cite{liao_equiformerv2_2024} & Desorption & 98 & 99  \\
    Eq2 (153M) & Dissociation & 81 & 90  \\
    Eq2 (153M) & Transfer & 68 & 84  \\
    \rowcolor{lightgray}
    Eq2 (31M)~\cite{liao_equiformerv2_2024} & Desorption & 98 & 97  \\
    \rowcolor{lightgray}
    Eq2 (31M) & Dissociation & 82 & 85  \\
    \rowcolor{lightgray}
    Eq2 (31M) & Transfer & 72 & 82  \\
    GemNet-OC~\cite{gasteiger2021gemnet} & Desorption & 98 & 95 \\
    GemNet-OC & Dissociation & 80 & 82 \\
    GemNet-OC & Transfer & 71 & 77 \\
    \rowcolor{lightgray}
    PaiNN~\cite{schutt_equivariant_2021} & Desorption & 98 & 95  \\
    \rowcolor{lightgray}
    PaiNN & Dissociation & 74 & 27  \\
    \rowcolor{lightgray}
    PaiNN & Transfer & 66 & 13  \\
    DimeNet++~\cite{klicpera2020fast} & Desorption & 95 & 94 \\
    DimeNet++ & Dissociation & 68 & 15 \\
    DimeNet++ & Transfer & 52 & 12 \\
    \end{tabular}
    \caption{Performance of Equiformer v2 (Eq2), GemNet-OC, PaiNN, and DimeNet++, trained on OC20 database, in predicting reaction (desorption, dissociation or transfer) barriers, tested on validation set based on OC20NEB. ``\% converged'' corresponds to the ratio of NEB calculations with converged forces, whereas ``\% success'' relates to the calculations in which the prediction of activation energy does not deviate from DFT values by more than 0.1~eV. The table is based on the results from Ref.~\cite{wander_cattsunami_2024}.} 
	\label{tab:neb_oc20}
\end{table}

\begin{figure}
	\centering
	\includegraphics[width=0.9\linewidth]{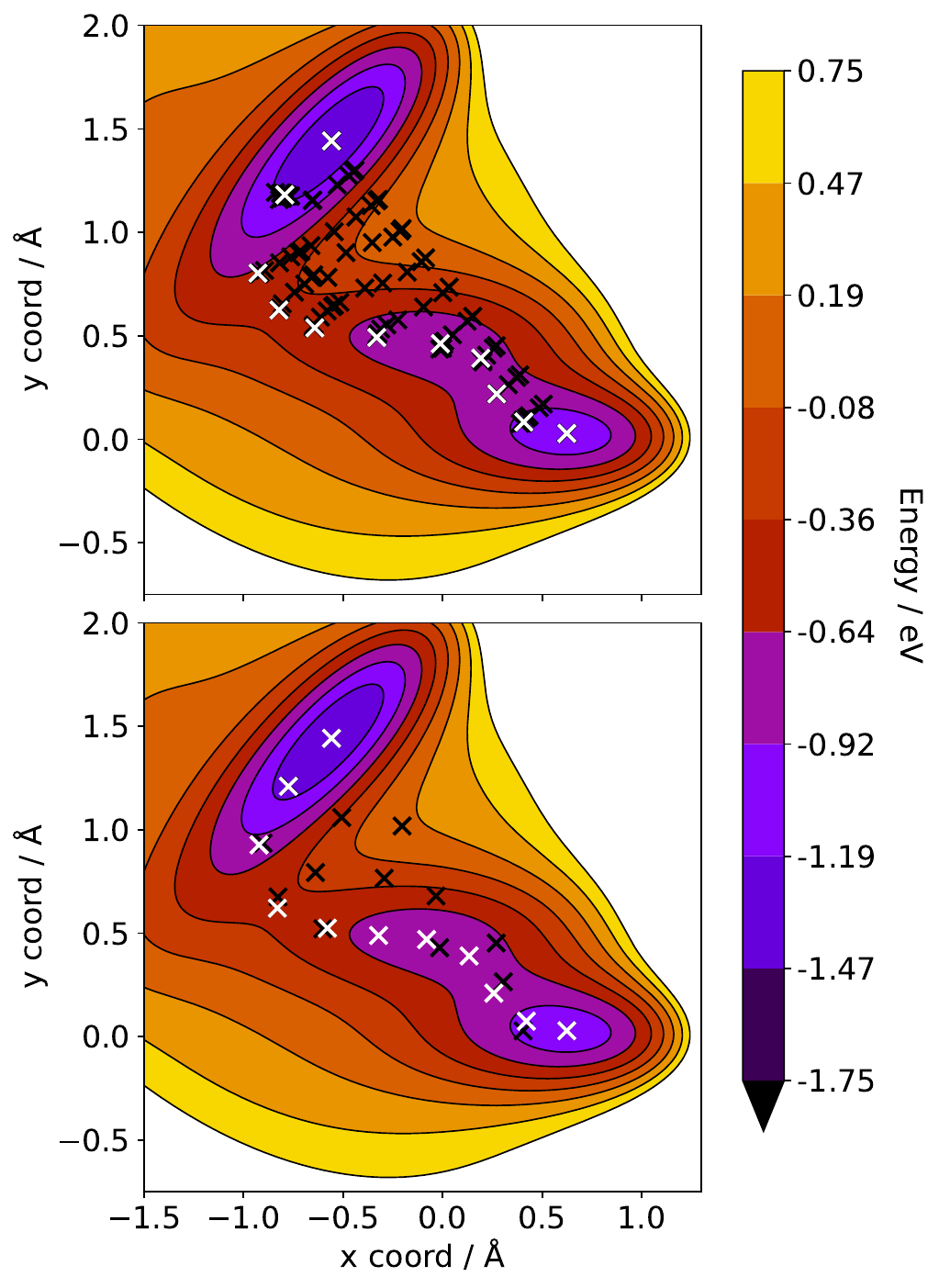}
	\caption{Minimum energy paths evaluated with climbing image NEB (above) and ML-NEB (below) based on the results from Ref.~\cite{PhysRevLett.122.156001} for the two-dimensional Müller-Brown PES. Black crosses correspond to all structures that were evaluated in the optimization process and white crosses represent the final path.}
	\label{fig:mlneb}
\end{figure}

To reduce the number of energy and force evaluations required during transition state search, Koistinen, Jónsson, and co-workers introduced GPR-aided NEB, which evaluates only the geometry from the highest uncertainty image of the predicted minimum energy path. The authors tested the algorithm on two cases: a 2D problem and a heptamer island on a (111) surface using Morse potentials ~\cite{koistinen_minimum_2016, koistinen_nudged_2017}. An improved GPR-based NEB (ML-NEB) was introduced by Garrido Torres and co-workers.~\cite{PhysRevLett.122.156001} Their algorithm adjusts the entire minimum energy path after every force evaluation, minimizing the number of evaluations needed to converge. The method was tested on Au diffusion on Al(111), Pt adatom diffusion on a stepped Pt surface, and a Pt heptamer island on Pt(111), demonstrating a remarkable reduction in force evaluations compared to established optimization algorithms. For the two-dimensional Müller-Brown potential, the search for the minimum energy path (F\textsubscript{max}=0.05~eV/\AA) using the climbing image NEB requires 286 evaluations (11 images), whereas with ML-NEB, it only requires 16 evaluations (Fig.~\ref{fig:mlneb}). Despite the success of the ML-NEB method in reducing the number of force evaluations needed to find the reaction barriers, it scales cubically with the number of atoms, thus it struggles for higher dimensional problems in which the model retraining step becomes a bottleneck that significantly increases the final evaluation time of NEB calculation.

\paragraph*{\textbf{Direct activation energy prediction}}

Apart from deriving activation energies and rates from transition states, both could in principle also be directly predicted without a prior search for the transition state structure. A {\ML} based direct prediction of activation energies in complex catalytic systems was proposed by Singh et al.~\cite{singh_predicting_2019} They employed a forward-search algorithm to select both linear and non-linear features and compared various ML techniques (linear regression, Gaussian process, random forest) for their effectiveness in predicting reaction rates. Focusing on the dehydrogenation and dissociation of N\textsubscript{2} and O\textsubscript{2} on surfaces, a polynomial-feature-based linear regression model performed the best, leading to the accuracy improvements of roughly 2 times over previous, single-parameter linear-regression-based models. Later, Komp et al. employed a more complex {\ML} model based on deep {\NN} to predict the full quantum reaction rate constant for one-dimensional reactive pathways using roughly 1.5 million training data points. The authors tested the model on the diffusion of H on Ni(100) and other non-surface reactions.~\cite{komp_machine_2020} 

Complementary experimental data can significantly enhance the predictions of kinetic properties. For example, Smith et al. proposed using experimental descriptor data combined with dimensionality reduction, principal component analysis, and {\NN} to predict reaction rates of the water-gas shift reaction using different catalysts and reaction conditions~\cite{smith_machine_2020}.

\paragraph*{\textbf{Open challenges.}}
The prediction of reaction barriers has reached a notable advancement with {\ML} techniques. However, an accurate and efficient direct reaction barrier prediction for molecules, materials, and surface chemistry has so far not been achieved. Promising methods, e.g. ML-NEB~\cite{PhysRevLett.122.156001}, have shown success in predicting reaction barriers of lower dimensional systems, however, more developments have to be made to provide such level of improvements robustly and consistently for higher-dimensional systems. NEB calculations typically demand substantial human intervention. Further development is necessary to create automated workflows that can efficiently identify barriers and transition states. Moreover, few comprehensive data sets, the OC20NEB database~\cite{wander_cattsunami_2024} being an exception, exist on barriers and transition paths which would facilitate development and testing of current and new methods.

\section{Muliscale kinetic simulations}
\label{sec:kinetic_simulations}

Kinetic simulations are essential for understanding surface reaction mechanisms and growth. Fundamental challenges in kinetic simulations include capturing processes that occur over vastly different time and length scales, accounting for a large chemical reaction space, and quantifying uncertainties and understanding how they propagate across scales. {\ML} can enhance large-scale kinetic simulations by integrating with mean-field microkinetic models, kinetic Monte Carlo (kMC) methods, and reaction networks. While microkinetic models are more computationally efficient, kMC simulations fully capture the reactive site dependence and fluctuations and reaction networks offer a framework for organizing complex reaction pathways and tracking the evolution of species over time.\cite{pineda_kinetic_2022} 

\paragraph*{\textbf{Mean-field microkinetic modeling}}

By assuming a uniform distribution of reactants and intermediates on the surface, mean-field microkinetic modeling simplifies the treatment of coverage-dependent effects and complex reaction networks. A mean-field model is the basis of the RMG-CAT software developed by Goldsmith and West.~\cite{goldsmith_automatic_2017} RMG-CAT employs graph theory supported by least-squares regression to simulate microkinetic mechanisms for heterogeneous catalysis and has been successfully applied to model the dry reforming of methane on a Ni surface. Tian and Rangarajan introduced a mean-field microkinetic approach that utilizes {\NN}s ({\NN}-MK). In their approach, rates of elementary steps in the fast diffusion limit are generated with the lattice MC simulations and then passed into an {\NN}-based model that maps coverages with reaction rate for the entire reaction network. This model was then used in the mean-field microkinetic model. They demonstrated the capabilities of the model by studying CO oxidation, reaching accuracy comparable to kMC simulations.~\cite{tian_machine-learned_2021, rangarajan_improving_2022-1}

\paragraph*{\textbf{Kinetic Monte-Carlo simulations}}

A more detailed approach to modeling kinetic processes is kMC, which enables capturing local fluctuations and spatial correlations. An early example of {\ML}-based kMC is the approach developed by Sastry et al.~\cite{sastry_genetic_2005} They used genetic programming to construct a symbolic regression for reaction barriers (given a small set of calculated barriers) to enable kMC simulations of the vacancy-assisted migration on an fcc Cu\textsubscript{x}Co\textsubscript{1-x} surface. Djurabekova et al. developed a simple {\NN} coupled with kMC to efficiently predict vacancy migration energies, reducing computational costs and enabling the exploration of kinetic pathways during Cu precipitation in Fe.~\cite{djurabekova_artificial_2007} Castin et al.~\cite{castin_calculation_2010} developed a similar approach of employing {\NN} model to predict vacancy migration energies based on NEB data, allowing acceleration of kMC simulations of thermal annealing in Fe-20\%Cr alloys, and later for precipitation studies in Fe-based alloys~\cite{castin_modeling_2011, castin_mobility_2012} and point-defect transition rates in FeCu alloys~\cite{messina_introducing_2017}. Castin et al. further extended this method with high-dimensional {\NN} {\MLIP}s to predict vacancy migration barriers in kMC, tested on FeCu and FeCr alloys~\cite{castin_improved_2017}.

Apart from using {\ML}-methods to predicting reaction rates and accelerating kMC simulations, it has also been proposed to entirely replace kMC with an {\ML}-based model. Chaffart et al. utilized {\NN}s to predict coefficients of stochastic partial differential equations as a function of substrate temperature and surface precursor fraction. The authors then combined the resulting method with continuum transport equations to predict epitaxial thin film evolution and growth of a gaseous molecular precursor.~\cite{chaffart_optimization_2018} Building on this approach, Kimaev et al. introduced a {\NN} that can completely replace the stochastic multiscale model, which included coupling kMC with partial differential equations to simulate thin film formation by chemical vapor deposition.~\cite{kimaev_nonlinear_2019, kimaev_multilevel_2020, kimaev_artificial_2020} Independently, Ding et al. developed an integrated multiscale recurrent {\NN}-based model for gas-phase transport profiles and microscopic surface dynamics using kMC and validated it for the plasma-enhanced atomic layer deposition of HfO\textsubscript{2} thin films.~\cite{ding_machine_2021} The improved efficiency of {\MLIP}s has recently enabled direct modeling of some kinetic processes. For instance, Zhou et al. employed a GPR-based active learning tool for constructing {\MLIP}s, named Flare~\cite{vandermause_--fly_2020}, to study reconstruction kinetics at the PdAu(111) surface induced by CO~\cite{zhou_dynamical_2022}. To efficiently learn the relation between reaction barriers and the output of the kMC model, Soleymanibrojeni et al. introduced an active learning process coupled with kMC to model solid-electrolyte interphase formation in Li-ion batteries, in which the initial dataset is constructed using a design-of-experiment approach, and a Gaussian process classification model.~\cite{soleymanibrojeni_active_2024}

\paragraph*{\textbf{Reaction networks}}
{\ML} models have shown great potential in uncovering the complexity of reaction networks that limit the modeling of many processes. Ulissi et al. presented a workflow that enables predicting the complex reaction pathway, employing GPR, which was applied to the syngas conversion on Rh(111).~\cite{ulissi_address_2017} 
Another approach was shown by Liu and co-workers who employed stochastic surface walking to construct global {\NN}-based {\PES}s (SSW-NN)~\cite{huang_material_2017} to study water gas shift reaction on the Cu(111) surface. In the study, DFT data was used containing 375,000 minima and more than 10,000 reaction pairs. The model enabled authors to find the lowest energy pathway for the entire reaction network.~\cite{kang_reaction_2021} This approach was later extended to an end-to-end framework for the activity prediction of heterogeneous catalytic systems (AI-Cat), in which two {\NN} models are used, the first one for predicting possible reaction patterns and the second one for predicting the reaction barrier and energy. The {\NN} models are employed in a Monte Carlo tree search to find the low-energy pathways of the reaction network. The authors applied AI-Cat to study the selectivity of glycerol hydrogenolysis on Cu surfaces.~\cite{kang_artificial_2022}

\paragraph*{\textbf{Open challenges.}}

Key challenges in kMC and microkinetic modeling are to ensure that the chemical reaction space (sites, elementary processes) adequately represents the system and that the underlying rate constants are valid for all relevant regimes. Rapid developments in data-driven approaches and global reaction exploration models are underway that will benefit the former. \cite{goldsmith_automatic_2017, ismail_graph-driven_2022} Robust and transferable {\ML} surrogates that can rapidly predict rate constants as a function of various conditions will dramatically benefit the latter.

The propagation of kMC and microkinetic models featuring processes with vastly different time scales is a challenge that continues to be an active research area. Approaches that aim to replace kMC with machine learning {\ML}-based models have shown promise. Additionally, sophisticated variable-coefficient differential equation solvers as well as robust uncertainty quantification and sensitivity analysis will be will be crucial for advancing this field.~\cite{gilkes_predicting_2024} 

Kinetic models not only couple to first principles data via input parameters, they also provide information for mass transport modelling of macroscopic surface models. {\ML} models will support not only support parameterization but also uncertainty quantification across different scales.

\section{Accurate dynamics at large time- and length scales}
\label{sec:dynamics_at_surfaces}

Simulating surface dynamics is challenging due to the high dimensionality, complex electronic structure of metallic surfaces, and the frequent requirement for ensemble averaging over thousands of molecular dynamics (MD) trajectories. Accurate modeling of dynamics at surfaces requires models that match the accuracy of first principles methods. However, employing ab initio MD for experimentally relevant systems is unfeasible in most cases, e.g. due to the long time-scales, and large system size, leading to high computational costs of ab initio methods. This challenge has driven the development of {\MLIP}s -- ML interatomic potentials trained to first principles accuracy.

\paragraph*{\textbf{Interatomic potentials}}

Simulations of dynamics on surfaces require highly efficient surrogate models for energies, force, and other properties. Fundamental to this are {\MLIP}s. We note that many {\MLIP}s are not solely dedicated to surface structure prediction and that their construction has recently been reviewed extensively in other areas of computational materials modeling.~\cite{deringer2019machine, mueller2020machine, mishin2021machine, fedik2022extending, zubatiuk2021development} Most existing methods to construct {\MLIP}s are based on {\NN}-potentials or Bayesian regression methods. {\NN}-potentials are unbiased and sufficiently flexible to learn from electronic structure data. They typically require in the range of $10^4$ training data points for high-dimensional {\PES} with tens to hundreds of degrees of freedom. Inference, i.e. evaluation of {\MLIP}s typically requires significantly more computational effort compared to empirical force field potentials. Beos et al. highlighted the advantages and disadvantages of {\NN}s over force fields by comparing a Behler-Parinello net to the ReaxFF~\cite{van2001reaxff} force field.~\cite{boes2016neural} While the {\NN} performed significantly better for bulk properties than ReaxFF, the {\NN} required significantly more (approximately ten times as much) training data compared to ReaxFF to obtain an accurate model for surface structures and adatom diffusion barriers. Bayesian regression methods are typically less data-hungry, but provide less flexibility and transferability.~\cite{hormann2019sample, todorovic2019bayesian, bartok2010gaussian}

Reactive gas-surface scattering simulations have long motivated the development of interatomic potentials. Early reactive {\PES} construction methods include the corrugation reducing procedure ~\cite{busnengo_surface_2005}, modified Shepard interpolation~\cite{ischtwan_molecular_1994, thompson_molecular_1997} or interpolation with permutation invariant polynomials~\cite{braams_permutationally_2009, bowman_high-dimensional_2011,jiang_enhancing_2012}. These methods enable low computational costs of MD simulations, however, they require using a frozen (static) surface approximation, excluding the explicit treatment of surface atom motion. Most recent dynamical simulations employ {\MLIP}s, which deliver ab initio accuracy and high efficiency while allowing the inclusion of all degrees of freedom in simulations. Therefore, {\MLIP}s have come to dominate surface dynamics research. Early {\MLIP}s for surface dynamics date back to 1995, when Blank et al. employed simple {\NN}s to construct PESs for CO adsorbed on a Ni(111) surface and H\textsubscript{2} on a Si(100) surface, utilizing the latter for quantum transition state theory rate calculations.~\cite{blank_neural_1995} A highly successful {\MLIP} based on Bayesian regression is GAP.~\cite{bartok2010gaussian} GAP allows interpolating the {\PES} by using GPR with a descriptor based on the local atomic density. GAP was originally applied to bulk crystals and was only later applied to problems in surface science.~\cite{timmermann2020iro, timmermann2021data, kloppenburg2023general} A milestone in the development of {\MLIP}s were the high-dimensional, atom-centered NNs introduced by Behler and Parrinello. Their method calculates the total energy by summing atomic energy contributions, with interatomic interactions described within a defined cutoff radius.~\cite{behler2007generalized} This atom-centered energy decomposition with finite cutoffs has since become a standard approach in most contemporary {\MLIP}s.
\begin{equation}
    E = \sum^{N_{\text{at}}}_{i=1}E_{i}=\sum^{N_{\text{at}}}_{i=1}NN_{i}
\end{equation}

A key challenge underpinning interatomic potentials is the selection of good atom-centered structural descriptors (or features). A descriptor is a mathematical representation of the local atomic environment, encoding structural and chemical information to enable a smooth {\ML} representation of energies, forces, and other properties as a function of the atomic positions and species. To be physically meaningful, a descriptor must remain invariant under the symmetry transformations of the system, such as rigid rotations, translations, and permutations. Correctly accounting for symmetries is crucial in {\MLIP}s, especially for surfaces, which often exhibit multiple symmetries. Behler et. al. were also among the first to apply {\NN}s in surface science. They used general symmetry functions based on atomic Fourier terms, replacing the inconvenient empirical functions presented by previous authors to describe the interaction between O\textsubscript{2} and a frozen Al(111) surface.~\cite{behler_representing_2007} Jiang, Guo, and colleagues introduced the permutation invariant polynomial {\NN} (PIP-NN), which uses permutation invariant polynomials as inputs, preserving both molecular permutation symmetry and surface translational symmetry. They applied this approach to simulate H\textsubscript{2} on frozen Cu(111) and Ag(111) surfaces.~\cite{jiang_permutation_2013, jiang_permutation_2014, jiang_six-dimensional_2014}

Methods based on high-dimensional NNs (HDNN) by Behler and Parrinello~\cite{behler2007generalized} allow the inclusion of surface degrees of freedom and form the most common type of {\MLIP} applied for dynamics at surfaces. Guo and co-workers were the first to use HDNNs, to investigate HCl scattering events on a dynamic Au(111) surface.~\cite{kolb_high-dimensional_2017, liu_constructing_2018} Gerrits employed the HDNN code RuNNer~\cite{behler2017first} to model H\textsubscript{2} reactions on a curved Pt crystal. This PES model, trained on data covering multiple Pt surfaces, allowed for larger unit cells without significant accuracy loss, making it suitable for simulating realistic crystal systems.~\cite{gerrits_accurate_2021} HDNNs were also used to study solid-liquid systems, in particular, for the dynamics of water at metal-based surfaces. Natarajan and Behler employed HDNN-based {\PES}s, to investigate water-copper interfaces for low-index Cu surfaces~\cite{natarajan_neural_2016-1} and stepped surfaces, including surface defects~\cite{natarajan_self-diffusion_2017}.

Jiang and co-workers developed the embedded atom {\NN} (EANN), which was shown to provide highly efficient force evaluations when compared to previous HDNN-based approaches and enabled accurate model construction with fewer data points.~\cite{zhang_embedded_2019} This method has been applied, for example, to simulate the scattering dynamics of NO on Ag(111)\cite{wang_rotationally_2023} or CO adsorption on Au(111)~\cite{hu_influence_2022,meng_vibrational_2022}. Later, a piece-wise EANN (PEANN) was proposed, in which the original Gaussian-type orbitals used for descriptors are replaced with piece-wise switching functions leading to significant improvement in efficiency when applied to dissociative chemisorption of CH\textsubscript{4} on flat Ir(111) and stepped Ir(332).~\cite{zhang_accelerating_2021} Notable improvement to EANN was introduced in recursively EANN (REANN), in which a message-passing scheme was implemented for orbital coefficients, leading to increased accuracy and transferability of the models. REANN was applied e.g. to model SiO\textsubscript{2} bulk or dissociative chemisorption of CO\textsubscript{2} at Ni(100).~\cite{zhang_physically_2021,zhang_reann_2022}

\paragraph*{\textbf{Equivariant interatomic potentials.}}
{\MLIP}s based on deep message-passing (MPNN) have been gaining significant traction. These models not only learn a target property but also the input descriptor in an end-to-end fashion. This promises a reduction of bias compared to hand-crafted descriptors. One of the first deep atom-centred message-passing {\NN} is the SchNet model.~\cite{schutt2017schnet, schutt2018schnet} A recent milestone is the advent of equivariant MPNNs, where the output transforms consistently with the input transformations, allowing them to capture the complex geometric relationships between atoms. Batzner et al. developed neural equivariant interatomic potentials (NequiP), an equivariant {\MLIP} based on MPNNs, that shows excellent accuracy in predicting energy and forces. It employs symmetry awareness by using E(3)-equivariant convolutions for interactions of geometric tensors. With this, the algorithm significantly reduces the amount of required data (up to 3 orders of magnitude) compared to other contemporary algorithms. The authors employed NequiP to study formate dehydrogenation on a Cu(110) surface.~\cite{batzner20223} Stark et al. explored the importance of equivariance by comparing invariant (SchNet~\cite{schutt2017schnet}) and equivariant (PaiNN~\cite{schutt_equivariant_2021}) MPNNs for H\textsubscript{2} dissociative adsorption on Cu surfaces, demonstrating that equivariant features enhance atomic environment descriptiveness, leading to more accurate models and smoother energy landscapes, while requiring fewer training data points.~\cite{stark2023}

\paragraph*{\textbf{Foundation models}}
Recently, several foundation models (also called universal {\MLIP}s), including MACE-MP-0,~\cite{batatia2023foundation} ANI/EFP,~\cite{haghiri2024ani} M3GNet-DIRECT,~\cite{qi2024robust} ALIGNN-FF,~\cite{choudhary2023unified} and CHGNet,~\cite{deng2023chgnet} have been introduced. These models are trained on a vast number of chemical species and data points with the ambition to provide a transferable prediction across a diverse range of systems. They have been successfully employed in modeling heterogeneous catalysis, or water/SiO\textsubscript{2} interface dynamics~\cite{batatia2023foundation}. Foundation models are pre-trained on large databases and often deliver qualitatively accurate predictions for tasks such as geometry optimization formation energy prediction.~\cite{yu2024systematic} Foundation models can suffer from softening (systematic underprediction of energies and forces), which can be overcome by retraining (or fine-tuning) on additional first-principles datasets (as few as 100 data points).~\cite{deng2405overcoming, focassio2024performance} Conversely, retraining foundation models may reduce their universality and lead to decreased performance on certain systems, including those present in the refinement dataset.~\cite{focassio2024performance} However, in many cases, compromising some degree of universality over accuracy for particular applications can be desirable. An example of such a case, for gas-surface dynamics (H\textsubscript{2} dissociation at Cu surfaces) was recently proposed by Radova et al. who introduced the MACE-freeze method,~\cite{radova_fine-tuning_2025} in which transfer learning with partially frozen model parameters is applied to the MACE-MP foundation model. The approach provided highly data-efficient MLIPs, in which similar accuracy to from-scratch-trained models is achieved with only 10-20\% data points. The authors also showed that training a light, linear model (ACEpotentials~\cite{witt_acepotentialsjl_2023}) based on data generated with the MACE-freeze model as ground truth can lead not only to better efficiency (17 times faster force evaluation as compared to the ``small'' MACE-MP model) but also to improved accuracy relative to from-scratch-trained linear models.

\paragraph*{\textbf{Data generation}}
Training of {\MLIP}s requires a large number of data points, the generation of which is computationally expensive, particularly due to the high computational cost associated with surface slab calculations. Adaptive sampling, also known as on-the-fly training, or active learning seeks to select the most informative data points to optimize the learning process with minimal data. Two strategies are particularly prevalent in surface science, uncertainty-based sampling and diversity sampling. In uncertainty-driven sampling, the model selects data points where its predictions are least confident, often focusing on points near decision boundaries or inherent uncertainties of a Bayesian model.~\cite{todorovic2019bayesian, xie2023uncertainty} Uncertainty-based sampling may involve balancing exploration and exploitation, as seen in Bayesian optimization methods. While leveraging uncertainties for exploration, they can also enhance exploitation e.g. by leveraging thermodynamic likelihood.~\cite{schwalbe-koda_differentiable_2021} Exploration can also be enhanced through additional techniques, e.g. by perturbing geometries.~\cite{van_der_oord_hyperactive_2023} Another uncertainty-based method is query-by-committee,\cite{seung1992query} where multiple models probe data points and new training points are selected where the models show the most disagreement. Query-by-committee was employed by Artrith and Behler in their development of {\NN} potentials for copper surfaces~\cite{artrith2012high}. Since then, its popularity has grown significantly, and we direct the interested reader to recent topical reviews.~\cite{kocer2022neural, tokita2023train}  Diversity sampling follows a different strategy by ensuring that selected data points are spread across the input space, preventing redundancy and ensuring a broad representation of the dataset. When a large amount of ab initio MD data is available, trajectory subsampling techniques can also be employed to extract the most informative data points from numerous trajectories.~\cite{zhang_bridging_2019} Interestingly, a combination of diversity sampling (clustering) and uncertainty-based sampling can significantly improve learning rates of {\MLIP}s compared to using exclusively one or the other method.~\cite{ghosh_active_2025}

\paragraph*{\textbf{Inclusion of long-range effects}}
Long-range dispersion interactions often constitute one of the dominant interactions at hybrid organic/inorganic interfaces.  For instance, the perylenetetracarboxylic dianhydride (PTCDA) molecule on Ag(111) is dominantly bonded by long-range interactions, despite also being anchored by peripheral oxygen atoms. ~\cite{hormann2020reproducibility, hofmann2021first} These interactions are non-local by nature and to describe them within an {\MLIP} may require non-local descriptors with larger cut-offs than otherwise needed. This may render {\ML}-models inefficient. Using SchNet, Westermayr et al. developed a long-range dispersion-inclusive {\MLIP} that facilitates structure search and geometry optimization of organic/inorganic interfaces.~\cite{westermayr2022long} They used two {\NN}s, one to learn the vdW-free interaction energy and one to learn the Hirshfeld volume ratios, allowing the calculation of vdW interactions as a correction to the vdW-free {\NN}.  Around the same time, Piquemal~\cite{poier2022accurate} and Caro~\cite{muhli2021machine} introduced similar approaches. Other long-range separated {\MLIP}s include the Long-Distance Equivariant (LODE) method,~\cite{cheng2019ab, huguenin2023physics} which employs local descriptors to represent Coulombic and other asymptotically decaying potentials around atoms, and the Latent Ewald Summation (LES)~\cite{cheng2024latent}, designed specifically to address long-range interactions in atomistic systems. Long-range electrostatics, in addition to long-range dispersion, have been included in the HDNN models by Behler~\cite{ko_fourth-generation_2021} and the SpookyNet model~\cite{unke_spookynet_2021}.

\paragraph*{\textbf{Benchmarking potential accuracy}}

The accuracy of {\MLIP}s is often assessed using standardized datasets such as MD17,~\cite{chmiela2017machine} QM9,~\cite{ramakrishnan2014quantum} OC20,~\cite{chanussot2021open} and OC22~\cite{tran2022open}.  The MD17 database contains MD trajectories for ten small organic molecules. The QM9 database contains relaxed geometries of 134,000 small, stable organic molecules composed of C, H, O, N, and F, for which geometric, energetic, electronic, and thermodynamic properties have been computed. Table \ref{tab:MLIP_accuracy} presents mean absolute errors for predicted energies. Almost all published {\MLIP} models have been assessed on MD17 or QM9 datasets, which both cover only organic molecules and are not directly relevant to surface science problems. More applicable to surface science is the open catalyst (OC) project.~\cite{chanussot2021open, tran2022open} The OC20 dataset comprises over 264 million data points, featuring relaxed and unrelaxed structures, adsorption energies, and atomic forces for various catalyst-adsorbate interactions. OC22 further extends OC20 with 9,8 million data points of complex reaction pathways and dynamic simulations, adding temporal data and focusing on reaction kinetics, making it a valuable resource for studying atomic-scale catalytic processes. Table \ref{tab:MLIP_accuracy} provides a comprehensive summary of literature benchmarks on these datasets. In the case of the OC20/22 datasets we focus on two different tasks highly relevant to surface science: (A) The first benchmark targets the prediction of the total energy, as calculated by DFT, for a given structure -- structure to energy and forces (S2EF). (B) The second benchmark targets the prediction of the relaxed DFT total energy for a given initial structure -- initial structure to relaxed structure (IS2RS).

\begin{table}[]
    \centering
    \setlength{\tabcolsep}{0pt}
    \begin{tabularx}{\linewidth}{Y|C{1.4cm}|C{1.4cm}|C{1.4cm}|C{1.4cm}}
     & \multicolumn{4}{c}{\textbf{MAE / meV}} \\
     & \textbf{MD17} & \textbf{QM9} & \textbf{OC22 S2EF} & \textbf{OC20 IS2RE} \\
    \hline
    DimeNet~\cite{klicpera2020fast} & \gradientcell{5.0}{2.0}{21.7} & \gradientcell{6.0}{4.1}{25.0} & \gradientcell{570.0}{263.0}{7924.0} & \gradientcell{661.3}{316.0}{851.0} \\
    PaiNN~\cite{schutt_equivariant_2021} & \gradientcell{14.3}{2.0}{21.7} & \gradientcell{5.8}{4.1}{25.0} & \gradientcell{2630.0}{263.0}{7924.0} & \gradientcell{743.0}{316.0}{851.0} \\
    SchNet~\cite{schutt2017schnet} & \gradientcell{6.9}{2.0}{21.7} & \gradientcell{14.0}{4.1}{25.0} & \gradientcell{7924.8}{263.0}{7924.0} & \gradientcell{703.9}{316.0}{851.0} \\
    EquiformerV2~\cite{liao_equiformerv2_2024} & & \gradientcell{6.17}{4.1}{25.0} & \gradientcell{659.8}{263.0}{7924.0} & \gradientcell{316.0}{316.0}{851.0} \\
    SpinConv~\cite{shuaibi2021rotation} & & \gradientcell{12.0}{4.1}{25.0} & \gradientcell{1944.0}{263.0}{7924.0} & \gradientcell{437.0}{316.0}{851.0} \\
    GemNet-dT~\cite{gasteiger2021gemnet} & & & \gradientcell{1271.3}{263.0}{7924.0} & \gradientcell{400.0}{316.0}{851.0} \\
    GemNet-OC~\cite{gasteiger2021gemnet} & & & \gradientcell{828.7}{263.0}{7924.0} & \gradientcell{344.0}{316.0}{851.0} \\
    NequIP~\cite{batzner20223} & \gradientcell{4.2}{2.0}{21.7} & & & \gradientcell{736.0}{316.0}{851.0} \\
    Equiformer~\cite{liao2022equiformer} & & \gradientcell{6.59}{4.7}{22.0} & & \gradientcell{603.0}{316}{851.0} \\
    Faenet~\cite{duval2023faenet} & & \gradientcell{6.79}{4.7}{22.0} & & \gradientcell{551.0}{316.0}{851.0} \\
    SEGNN~\cite{yu2022graph} & & \gradientcell{15.0}{4.7}{22.0} & & \gradientcell{679.0}{316.0}{851.0} \\
    SphereNet~\cite{coors2018spherenet} & & \gradientcell{6.0}{4.7}{22.0} & & \gradientcell{637.8}{316.0}{851.0} \\
    CGCNN~\cite{xie2018crystal} & & & & \gradientcell{851.0}{316.0}{851.0} \\
    EAA~\cite{chang2023molecular} & \gradientcell{5.9}{0}{21.7} & \gradientcell{12.0}{4.1}{25.0} & & \\
    GM-sNN~\cite{zaverkin2020gaussian} & \gradientcell{7.1}{0}{21.7} & \gradientcell{11.7}{4.1}{25.0} & & \\
    PhysNet~\cite{unke2019physnet} & \gradientcell{5.3}{2.0}{21.7} & \gradientcell{8.2}{4.1}{25.0} & & \\
    TorchMD-NET~\cite{tholke2022torchmd} & \gradientcell{3.6}{2.0}{21.7} & \gradientcell{6.2}{4.1}{22.0} & & \\
    ANI~\cite{smith2017ani} & \gradientcell{21.7}{2.0}{21.7} & & & \\
    ACE~\cite{kovacs2021linear} & \gradientcell{2.0}{2.0}{21.7} & & & \\
    EANN~\cite{zhang_embedded_2019} & \gradientcell{6.4}{2.0}{21.7} &  &  &  \\
    FCHL~\cite{faber2018alchemical} & \gradientcell{4.6}{2.0}{21.7} &  &  &  \\
    GAP~\cite{bartok2010gaussian} & \gradientcell{16.1}{2.0}{21.7} &  &  &  \\
    NewtonNet~\cite{haghighatlari2022newtonnet} & \gradientcell{5.0}{2.0}{21.7} &  &  &  \\
    REANN~\cite{zhang_reann_2022} & \gradientcell{4.4}{2.0}{21.7} &  &  &  \\
    sGDML~\cite{chmiela2018towards} & \gradientcell{5.0}{2.0}{21.7} &  &  &  \\
    So3krates~\cite{frank2022so3krates} & \gradientcell{4.3}{2.0}{21.7} & & & \\
    Allegro~\cite{musaelian2023learning} & & \gradientcell{4.7}{4.1}{25.0} & & \\
    Cormorant~\cite{anderson2019cormorant} &  & \gradientcell{22.0}{4.1}{25.0} &  &  \\
    EGNN~\cite{satorras2021n} &  & \gradientcell{11.0}{4.1}{25.0} &  &  \\
    EQGAT~\cite{le2022equivariant} &  & \gradientcell{25.0}{4.1}{25.0} &  &  \\
    LieConv~\cite{finzi2020generalizing} &  & \gradientcell{19.0}{4.1}{25.0} &  &  \\
    MACE~\cite{batatia2022mace} & & \gradientcell{4.1}{4.1}{25.0} & & \\
    NMP~\cite{gilmer2017neural} &  & \gradientcell{20.0}{4.1}{25.0} &  &  \\
    \end{tabularx}
    \caption{MAE in meV for the prediction of energies (and geometry optimizations in case of OC20 IS2RE); Blue and red colors indicate small and large MAEs respectively; {\MLIP}s trained and tested on MD17 (1,000 training data points), QM9 (110,000 training data points), OC20 (460,328 training data points), and OC22 (8,225,293 training data points); MAEs on OC20/22 for out-of-domain set; MAEs are taken from~\cite{lo2023training, liao_equiformerv2_2024, tran2022open, kim2024geotmi, duval2023faenet, ramlaoui2024improving, shoghi2024distribution, chang2023molecular, shoghi2023molecules, batzner20223, zhang_reann_2022, kovacs2021linear, frank2022so3krates, tholke2022torchmd, liao2022equiformer, shuaibi2021rotation}; A more detailed table with MAEs can be found in the Supporting Information.}
    \label{tab:MLIP_accuracy}
\end{table}

\paragraph*{\textbf{Benchmarking speed of prediction of potentials}}

\begin{figure}
	\centering
	\includegraphics[width=1.0\linewidth]{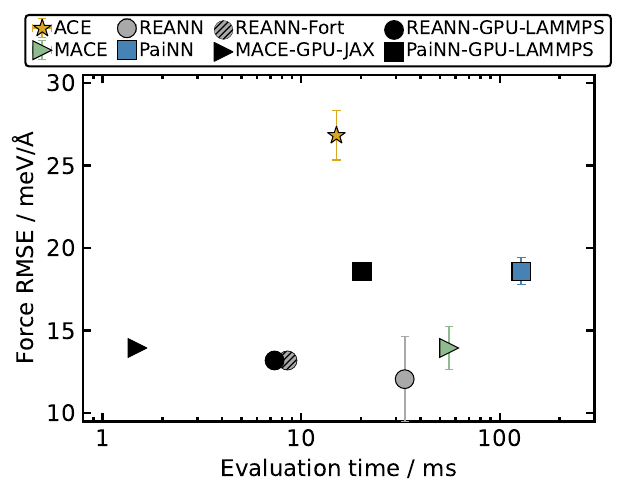}
    \caption{Force test RMSEs in meV/$\textrm{\AA}$ plotted against force evaluation time in milliseconds from Ref.~\cite{stark_benchmarking_2024} obtained with models based on different architectures, such as PaiNN (squares), REANN (circles), MACE (triangles), and ACE (stars). Data points and error bars correspond to the RMSE average and standard deviation over 5-fold cross-validation splits. The striped circle corresponds to the model obtained with the newer REANN version, which uses a Fortran-based neighbor list calculator. Black markers indicate models evaluated using GPU architecture. The CPU evaluation times were calculated using a single AMD EPYC 7742 (Rome) 2.25~GHz CPU processor core. GPU evaluation times were obtained with NVIDIA V100 GPU.}
	\label{fig:error_vs_time}
\end{figure}

Often the primary measure of the quality of {\MLIP}s is the prediction accuracy, which is usually determined by the mean absolute error (MAE) and the root mean square error (RSME) on a test set. However, time-to-solution can be equally important in the context of the need for accurate statistical sampling of nonequilibrium gas-surface dynamics. Stark et al. benchmarked {\MLIP}s such as ACE, MACE, REANN, and PaiNN~\cite{stark_benchmarking_2024}, evaluating accuracy via statistical ensemble averaging and comparing evaluation speeds on CPU and GPU architectures (Fig.~\ref{fig:error_vs_time}). MACE and REANN models achieved good accuracy and efficiency within CPU architectures that are most suitable for massively parallel multi-trajectory dynamics. MACE provided a superior performance within GPU architectures. As a linear model, ACE is the fastest on CPU architectures, but providing training data to ensure consistent accuracy is more challenging than for {\MLIP}s.

\paragraph*{\textbf{Open challenges.}}

What is evident from the overview in Table \ref{tab:MLIP_accuracy}, is the existence of a benchmarking gap, as most interatomic potentials are commonly evaluated on molecular datasets rather than surface structures. Notably, various potentials, such as ACE, MACE, or REANN are routinely applied in surface dynamics but were not yet benchmarked on the OC20/22 datasets. Moreover, there exists an accuracy gap: For example, the OC20/22 datasets were determined using GGA(+U) level of theory,~\cite{chanussot2021open, tran2022open} while the molecular database QM9 uses hybrid-DFT (B3LYP) level of theory~\cite{ramakrishnan2014quantum}. Surfaces and interfaces are governed by mechanisms such as hybridization, charge transfer, Pauli repulsion, vdW interactions, level alignment, and surface-mediated electronic states~\cite{hofmann2021first}, often requiring beyond-GGA accuracy, as exemplified by the CO on metals puzzle~\cite{gajdovs2005co, feibelman2001co}. To further complicate things, relevant systems frequently comprise hundreds or thousands of atoms, limiting the ability to use a suitable level of theory. These challenges make generating the underlying data difficult and computationally expensive and can introduce greater intrinsic uncertainty due to varying convergence thresholds. 

\section*{Excited states and nonadiabatic dynamics\label{sec:nonadiabatic}}

\begin{figure*}
	\centering
	\includegraphics[width=0.95\linewidth]{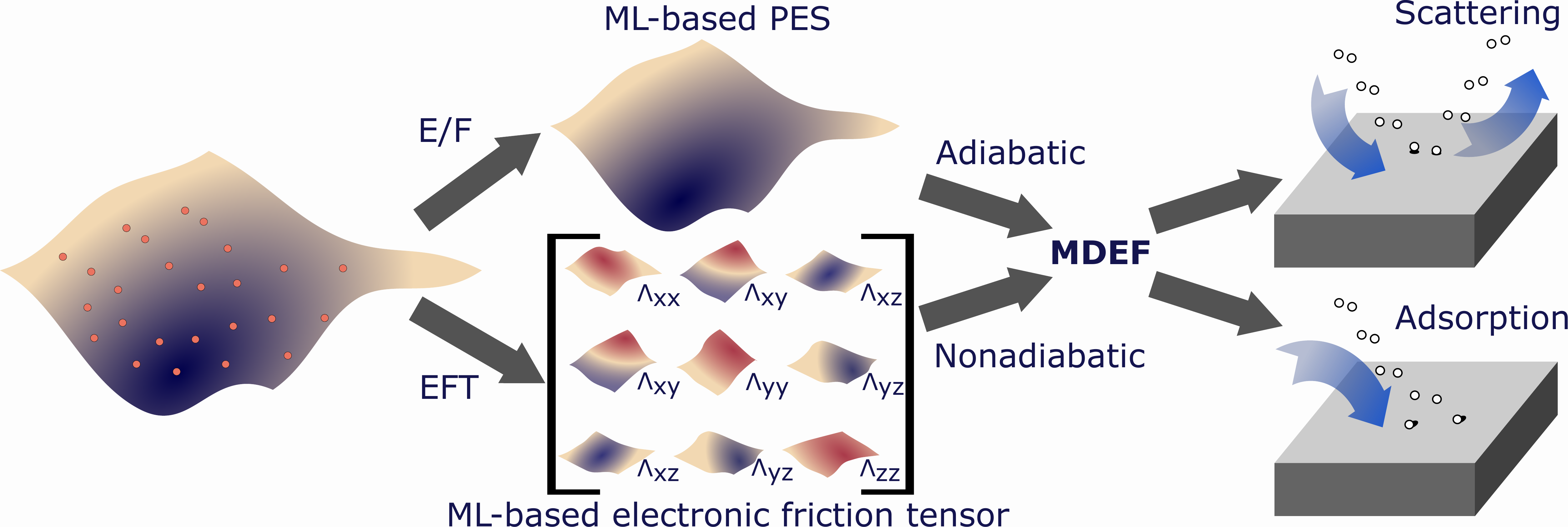}
	\caption{Exploration of nonadiabatic effects at metal surfaces using MDEF. {\MLIP}s can be combined with electronic friction ML models within the MDEF to accelerate MD simulations beyond the adiabatic approximation.}
	\label{fig:ml_md}
\end{figure*}

In most cases, the Born-Oppenheimer approximation is valid for describing the dynamics at surfaces, however, especially when considering dynamics at metals or low-bandgap semiconductor surfaces, electronic excitations and ensuing nonadiabatic effects cannot be ignored. Examples of approaches that can be used to simulate nonadiabatic electron-nuclear coupling effects in MD include Ehrenfest dynamics, a range of trajectory surface hopping methods~\cite{tully_molecular_1990,Shenvi2009}, and MD with electronic friction (MDEF)~\cite{headgordon_molecular_1995}. MDEF has been commonly used in the study of reactive scattering and light-driven dynamics of small molecules at metal surfaces. In this method, the dynamics are propagated on a single ground state {\PES}. Atoms experience additional forces due to electronic friction and a random white noise term within a Langevin dynamics framework. 
Two approaches have been proposed to calculate the electronic friction tensor (EFT) from first principles: The local density friction approximation (LDFA)~\cite{juaristi_role_2008} and first-order response theory based on DFT (otherwise known as orbital dependent friction or ODF)~\cite{askerka_role_2016,maurer_ab_2016}. In LDFA, the EFT is reduced to scalar friction values based on a bare surface electron density. ODF is calculated from Kohn-Sham DFT through time-dependent perturbation theory and provides a full EFT.

Within a static surface approximation, LDFA can be efficiently evaluated from first principles, requiring only a single calculation of the surface electron density. Alducin, Juaristi, and co-workers published a series of studies on light-driven surface reactions, incorporating {\NN}-based {\MLIP}s in combination with LDFA friction, e.g. to study laser-driven desorption~\cite{lindner_femtosecond_2023,muzas_multicoverage_2024}. However, especially at high temperatures, the inclusion of surface degrees of freedom may be crucial in simulating processes at surfaces, which can benefit from the use of flexible and efficient surrogate models of the EFT in addition to the {\MLIP} that describes the {\PES}. 
Alducin, Juaristi, and co-workers addressed this in later studies by employing a simplified density model based on exponentially decaying functions, that allows fast and accurate predictions of electronic friction within LDFA as a function of surface atom motion~\cite{serrano_jimenez_photoinduced_2021,muzas_absence_2022,zugec_understanding_2024}. 

Utilizing {\ML} techniques to represent the EFT is crucial within first-order perturbation theory, (also called ODF) due to the high cost of linear response calculations in DFT. Spiering, Meyer and co-workers introduced a symmetry adapted six-dimensional {\NN}-based EFT model for H\textsubscript{2} and D\textsubscript{2} on Cu(111) and for N\textsubscript{2} on Ru(0001) surface~\cite{spiering_testing_2018, spiering_orbital-dependent_2019}. Zhang, Maurer, and co-workers created a {\NN}-based EFT model that accounts for the covariance properties of the EFT with respect to the surface symmetry using a simple mapping scheme and used it to study the reactive scattering of H\textsubscript{2} on Ag(111) surface~\cite{zhang_hot-electron_2019, maurer_hot_2019-1}. The authors further improved the model by preserving the positive semidefiniteness, directional property, and correct symmetry-equivariance of EFT and tested it on the same example of H\textsubscript{2} dynamics at Ag(111) surface~\cite{zhang_symmetry-adapted_2020}. The model was also employed to study NO dynamics on Au(111).~\cite{box_determining_2020} Recently, Sachs et al.~\cite{sachs_equivariant_2024} introduced an Atomic-Cluster-Expansion-based EFT model (ACEfriction), utilizing equivariant representations of tensor-valued functions that satisfy all the symmetries of EFT by construction, allowing highly accurate and efficient prediction of friction and diffusion tensors. The construction of ACEfriction provides for size-transferability by enabling the prediction of EFTs of multiple adsorbates and larger friction tensors. The model was tested on NO/Au(111) system. It was also applied by Box et al. in the context of H atom scattering on Pt~\cite{box_room_2024}.

Nonadiabatic effects are also explored with other techniques. Several {\ML} models were constructed for effective surrogate Hamiltonians coupled most commonly with trajectory surface hopping dynamics methods~\cite{dral_nonadiabatic_2018,zhang_doping-induced_2021,wang_interpolating_2021,prezhdo_modeling_2021,zhou_deep_2022,akimov_extending_2021,wang_all-atom_2022,shakiba_machine-learned_2024}. For example, Liu~et~al. combined an ML-based Hamiltonian surrogate and force fields with decoherence-induced surface hopping to study defects in MoS\textsubscript{2}~\cite{liu_breaking_2024}. Meng~et~al. used {\ML} for predicting excited states based on constrained {\DFT} to construct an effective diabatic Hamiltonian, which was propagated with independent electron surface hopping dynamics. This was used to study nonadiabatic dynamics of CO scattering on Au(111) system.~\cite{meng_first-principles_2024} ML surrogate models accelerate and crucially enable quantum and mixed-quantum-classical dynamics in gas-surface dynamics, which makes this an exciting application area of ML in the coming years.

\paragraph*{\textbf{Open challenges.}}

Most applications on excited state and nonadiabatic dynamics on surfaces have so far focussed on dynamics in the classical path, mean-field approximation or molecular dynamics with electronic friction. These methods have uncontrolled and insufficiently assessed errors and are limited to situations of weak nonadiabaticity and strong adsorbate-substrate hybridisation.~\cite{gardner_assessing_2023} To enable the application of more robust mixed quantum-classical simulation methods to high-dimensional surface systems, ML electronic structure surrogates (as described in \hyperref[sec:nonadiabatic]{Excited states and nonadiabatic dynamics}\ref{sec:electronic_property}) will play a crucial role. These models will need to be able to cope with little data of suboptimal accuracy. Accurate and scalable first principles reference calculations of excited-state properties of surfaces, for example based on Many Body Perturbation Theory, are not routinely available for data generation. Equally important will be the development of improved dynamics methods that can efficiently capture nonadiabatic electron-phonon coupling, but also quantum decoherence and electron-electron scattering effects, which cannot be neglected in surface systems with high electronic DOS.

\section{\label{sec:conclusions}State of the Field and Future Perspective}

Data-driven and {\ML} methods offer powerful tools to accelerate surface structure determination and energy and force prediction in surface dynamics simulations. {\ML} approaches are used to accelerate geometry optimizations, enabling precise determinations of molecular adlayers and surface clusters in systems with many degrees of freedom. They also aid in determining energy barriers and transition states, with {\ML}-accelerated NEB methods as a prominent example. {\ML} is increasingly used to enhance or even replace kMC methods. Early approaches to improve the efficiency of surface dynamics relied on corrugation-reduction or the modified Shepard interpolation. Recently machine-learned {\PES} were used to attain cost-efficient energies and forces for MD simulations. Beyond the Born-Oppenheimer approximation, {\ML} approaches are applied to learn electronic friction or excited-state surrogate Hamiltonians.

However, {\ML} and data-driven methods in surface science often focus on simple or idealized systems, neglecting surface reconstructions, assuming low temperatures and ultra-high vacuum, focusing on single atoms or small molecules, or imposing commensurability. Highly relevant challenges, such as modeling incommensurate structures, kinetics and growth, high-pressure and high-humidity systems, solid-liquid interfaces with charged ions, electrochemical potentials and conditions, and light-matter interactions, push the limits of current approaches. Bridging this complexity gap requires efficient, massively scalable, and easily deployable {\ML} tools trained on large, well-curated, and accurate datasets built on a synthesis between computational and experimental data. Hereafter, we will summarize the most important challenges to the application of ML methods in surface science.

\paragraph*{\textbf{Benchmarking gap.}}

There is a pressing need for more comprehensive benchmarking of {\MLIP}s for surface simulations. As highlighted in Table \ref{tab:MLIP_accuracy}, recent advancements in dynamic model development are promising, but comprehensive benchmarking of widely used {\MLIP}s on specific surface science datasets such as OC20/22, remains insufficient. Moreover, benchmarking on databases containing barriers (OC20NEB~\cite{wander_cattsunami_2024}) and thermodynamic properties remains largely unaddressed. Addressing this gap calls for a community-wide effort to test a broader range of models on existing surface science datasets while also creating more and more challenging datasets. These should extend beyond equilibrium geometries and transition state data, capturing the complexity of gas-surface dynamics and other non-equilibrium dynamical phenomena. Establishing a standardized benchmark dataset for gas-surface dynamics could provide a valuable foundation for the assessment of {\MLIP}s.

\paragraph*{\textbf{Accuracy gap.}}

Most current datasets in surface science are derived from DFT, predominantly at GGA level of theory. While GGA is widely used, it is often inadequate to accurately capture key surface properties, particularly energy barriers and adsorption energies. Datasets based on hybrid DFT or more advanced exchange-correlation energy descriptions and beyond-DFT methods remain scarce, and their reliability for surface science problems is not fully understood.~\cite{kroes2021computational} Even more critically, there is a significant lack of high-quality experimental data on structure, stability, and kinetics to guide or validate theoretical models.~\cite{maurer2016adsorption} Relying solely on learning from GGA-DFT data will not fundamentally address the challenges in the electronic structure description of surfaces and interfaces.

Computational surface science data often contains more intrinsic noise and uncertainty compared to molecular data, which needs to be considered when applying {\ML} methods. This is because surfaces are inherently more complex and require larger system sizes than molecules or bulk materials.~\cite{hofmann2021first} The difficulty in imposing tight convergence criteria results in the absence of a definitive "gold standard" for accurate first-principles surface computations. Depending on the system, different computational approaches may be suitable.~\cite{hofmann2021first} Exchange-correlation functionals differ widely in their predictive capabilities and even for the same density functional approximation, different codes provide different results due to numerical approximations.~\cite{lejaeghere2016reproducibility}

\paragraph*{\textbf{Data gap.}}

Many computational approaches in surface science are data-hungry, often requiring thousands of data points for accurate predictions. Computational data generation is costly due to the high demands of quantum mechanical calculations for large systems, particularly those with long-range interactions such as incommensurate structures or surface reconstructions. Experimental data generation often requires highly controlled environments and time-consuming sample preparation. Efforts to address this include efficient training algorithms such as optimal selection and active learning, which rely on estimators for information gain. A common estimator is the model uncertainty, which may, however, often fail to align with true errors, leading to overconfidence in poorly trained regions and a failure to generalize effectively. In highly regularized equivariant MPNNs, uncertainty estimation methods like bootstrapping or committees are less effective.~\cite{gasteiger2020fast} Addressing these shortcomings requires improvements in uncertainty quantification techniques, better validation protocols, transfer learning, and the incorporation of multi-fidelity approaches that combine high-accuracy data with less expensive, approximate calculations.

Transfer learning has gained increasing attention, particularly with the rise of foundation models (see \hyperref[sec:dynamics_at_surfaces]{Accurate dynamics at large time- and length scales}). While research has primarily focused on {\MLIP}s, there remains a significant gap in developing reliable methods to assess the transferability of {\ML} models across systems. Ensuring transferability between small- and large-scale systems, or across different surfaces and compositions, is crucial. Additionally, generating validation data for large systems will be necessary to confirm this transferability. Little research has been conducted so far on transfer learning for the direct prediction of kinetic properties, barriers, or other properties beyond energies and forces.


The scarcity of high-fidelity computational and experimental data underscores the urgent need for data synthesis methods and multi-fidelity learning, such as multi-head {\NN}s and transfer learning.\cite{cui_multi-fidelity_2024, elena_machine_2025, radova_fine-tuning_2025} These methods can synthesize information across different levels of theory—combining data from lower-accuracy methods (e.g., GGA-DFT) with more accurate beyond-DFT calculations, such as hybrid DFT, GW, or random phase approximation, and experimental observations. Data augmentation methods, where experimental data is supplemented with synthetically generated datasets, have shown promise for improving the analysis of scattering experiments.~\cite{anker2023machine} Integrated approaches will help to bridge the gaps between theoretical approximations and experiments, for instance enhancing the modeling of transition barriers, reaction rates, phase diagrams, and spectroscopy. They may also be able to optimize experiments and simulations by guiding resource allocation and improving data efficiency.

\paragraph*{\textbf{Open and FAIR data.}}

Computational surface science generates vast amounts of data, but only a fraction is typically relevant to specific tasks. This data must be publicly available and well-curated, in compliance with the FAIR data principles~\cite{wilkinson2016fair}. Table \ref{tab:databases} lists common databases that feature surface science data. Most databases are not exclusive to surface science and contain data from the wider field of computational chemistry/physics. Establishing large, consistent datasets is crucial for advancing {\ML} in computational surface science. To foster reproducibility, knowledge transfer, and data reuse, we urge the community to make well-curated computational data publicly accessible.

\begin{table}[h]
    \centering
    \begin{tabularx}{\linewidth}{Yl}
        AFLOW~\cite{curtarolo2012aflow} & \href{https://www.aflowlib.org/}{www.aflowlib.org/}\\\hline
        AiiDA~\cite{pizzi2016aiida} & \href{https://www.aiida.net/}{www.aiida.net}\\\hline
        ASR~\cite{gjerding2021atomic} & \href{https://asr.readthedocs.io/en/latest/}{www.asr.readthedocs.io}\\\hline
        Catalysis hub~\cite{winther2019catalysis} & \href{https://www.catalysis-hub.org/}{www.catalysis-hub.org}\\\hline
        Catalyst property database~\cite{chemcatbio} & \href{https://cpd.chemcatbio.org/}{www.cpd.chemcatbio.org}\\\hline
        Materials cloud~\cite{talirz2020materials} & \href{https://www.materialscloud.org}{www.materialscloud.org}\\\hline
        Materials project~\cite{jain2013commentary} & \href{https://materialsproject.org}{www.materialsproject.org}\\\hline
        MAX~\cite{max} & \href{http://www.max-centre.eu}{www.max-centre.eu}\\\hline
        NOMAD~\cite{draxl2019nomad} & \href{https://nomad-lab.eu/nomad-lab/}{www.nomad-lab.eu}\\\hline
    \end{tabularx}
    \caption{List of common databases for computational materials data as well as automated data generation tools}
    \label{tab:databases}
\end{table}

\paragraph*{\textbf{Efficient and accurate inference.}}

A notable challenge is the improvement of inference performance. Many current {\ML} models are, while powerful and accurate, too computationally expensive for high-throughput predictions of properties or long-time-scale and large-system-size dynamics simulations in surface science. This is particularly important for nonequilibrium dynamics at surfaces that require ensemble averaging over tens of thousands of trajectories (e.g. to evaluate quantum-state-resolved reactive scattering probabilities). 
New developments in the accurate and efficient description of atomic environments and in {\MLIP} architectures continue to improve accuracy and transferability, while reducing the computational cost of model evaluation, as shown in Fig.~\ref{fig:error_vs_time}. GPU architectures and workflows can further reduce computational costs, however, the current bottleneck in deploying or renewing GPU compute architectures should also encourage us to continue to seek faster prediction on CPU architectures and models that benefit from just-in-time compilation.

\paragraph*{\textbf{Scalability and deployability}} are equally important. Efficiency goes beyond simply reducing inference time, it involves creating scalable and automated workflows that are adaptable to diverse and heterogeneous computing architectures. For complex simulations of nonequilibrium dynamics at surfaces—such as gas-surface interactions, atomic layer deposition, chemical vapor deposition, or rare event sampling—an effective ML approach must support high-throughput ensemble sampling. Achieving this requires solutions that balance computational loads across GPUs and CPUs efficiently. ML frameworks must support batched evaluations to maximize GPU utilization while simultaneously enabling parallel task farming for ensemble dynamics propagation. This dual focus ensures scalability in handling both individual trajectory evaluations and the broader orchestration of ensemble simulations. Furthermore, workflow integration is key. Effective ML deployments for surface science must seamlessly integrate data preprocessing, on-the-fly model evaluation, and post-processing while maintaining robustness and flexibility.\cite{gardner2022nqcdynamics, stark_benchmarking_2024} Moreover, it is essential to integrate {\ML} model predictions with downstream tasks, such as predicting electronic properties or Hamiltonians, as well as incorporating them into efficient bandstructure calculations.\cite{10.1063/5.0209742}

\paragraph*{\textbf{Machine learning and the future of surface characterization.}}

Multi-technique characterization is a central paradigm in surface science, encompassing various spectroscopic methods, diffraction techniques, and imaging approaches such as scanning probe microscopy. Despite their importance, there has been limited progress in leveraging {\ML} to bridge the gap between simulation and experiment in these techniques. This presents significant untapped potential for innovation and advancement as ML can use atomistic simulation data to complement highly integrated experimental measurement signals. Recent developments illustrate this promise, including the prediction of spectroscopic properties such as Raman and surface-enhanced Raman scattering measurements,~\cite{lussier2020deep} surface characterization using deep learning and infrared spectroscopy~\cite{yu2023surface}, and classification in X-ray absorption and emission spectroscopy~\cite{tetef2021unsupervised}. Efforts such as learning metastable phase diagrams~\cite{srinivasan2022machine}, reaction networks~\cite{ulissi2017address}, and catalytic activity from experimental data~\cite{smith2020machine} pave the way to learning end-to-end transition rates from experimental data. These efforts demand extensive experimental databases supported by automated labs, similar to the recently proposed OCx24 database~\cite{abed_open_2024}. 
Finally, efforts such as the application of reinforcement learning to automate scanning probe experiments~\cite{ramsauer2023autonomous} will benefit from models that are aware of hidden atomic-scale conformations~\cite{leinen_autonomous_2020} for example by being trained on complementary first-principles data. These examples underscore the vast opportunities for {\ML} to revolutionize surface science. 

Surface science, perhaps more than other fields, offers a unique opportunity to integrate highly controlled experiments aimed at uncovering fundamental physics with the power of {\ML} and data-driven approaches. Its distinct challenges, combined with its significant industrial relevance, continue to drive innovation in computational methods, fostering advancements with transformative impacts that extend well beyond the field itself.


\section*{acknowledgments}
The authors acknowledge financial support from the UKRI Future Leaders Fellowship program (MR/S016023/1, MR/X023109/1), a UKRI Horizon grant (ERC StG, EP/X014088/1), a Leverhulme Trust research project grant (RPG-2019-078), a UKRI Horizon grant (MSCA, EP/Y024923/1), and a UFO (Unkonventionelle Forschung) postdoctoral fellowship grant by the Austrian province of Styria.

\section*{Appendix 1: List of abbreviations and their descriptions}

\begin{table}[h!]
\centering
\begin{tabularx}{\linewidth}{Yp{6cm}}
\hline
\textbf{Abbreviation} & \textbf{Description} \\
\hline
CPU     & Central processing unit \\
DFT     & Density Functional Theory \\
DOS     & Density of states \\
EFT     & Electronic Friction Tensor \\
GA      & Genetic Algorithm \\
GGA     & Generalized Gradient Approximation \\
GPR     & Gaussian Process Regression \\
GPU     & Graphics processing unit \\
HDNN    & High-Dimensional Neural Network \\
kMC     & Kinetic Monte Carlo \\
LDFA    & Local Density Friction Approximation \\
MAE     & Mean Absolute Error \\
MD      & Molecular Dynamics \\
MDEF    & MD with Electronic Friction \\
ML      & Machine Learning \\
ML-NEB  & GPR-based NEB \\
MLIP    & Machine Learned Interatomic Potential \\
MPNN    & Deep Message-Passing Neural Network \\
NEB     & Nudged Elastic Band \\
NEXAFS  & Near edge X-ray absorption fine structure \\
NN      & Neural Network \\
ODF     & Orbital Dependent Friction \\
PES     & Potential Energy Surface \\
RMSE    & Root Mean Square Error \\
UPS     & Ultraviolet photoelectron spectroscopy \\
XPS     & X-ray photoelectron spectroscopy \\
\hline
\end{tabularx}
\caption{List of abbreviations and their descriptions.}
\label{tab:abbreviations}
\end{table}


\providecommand{\noopsort}[1]{}\providecommand{\singleletter}[1]{#1}%

\section*{Appendix 2: List of abbreviations and their descriptions}

\begin{table*}[]
  \centering
  \setlength{\tabcolsep}{0pt}
  \begin{tabularx}{\linewidth}{Y|C{4.4cm}|C{3.2cm}C{3.6cm}|C{3.6cm}}
   & \textbf{MD17} & \textbf{QM9} & \textbf{OC22} & \textbf{OC20 IS2RE} \\
  \hline
  DimeNet~\cite{klicpera2020fast} & 5.0,~\cite{batzner20223} 15.6~\cite{kovacs2021linear} &6.0,~\cite{liao2022equiformer} 
  6.3~\cite{tholke2022torchmd} & 570.0,~\cite{lo2023training} 2475.1~\cite{tran2022open} & 661.3,~\cite{shoghi2024distribution} 683.5~\cite{chanussot2021open}\\
  PaiNN~\cite{schutt2021equivariant} & 14.3,~\cite{kovacs2021linear} 4.6*~\cite{batatia2022mace} & 5.8~\cite{liao2022equiformer} & 2630.0~\cite{tran2022open} & 743.0~\cite{shoghi2024distribution} \\
  SchNet~\cite{schutt2017schnet} & 6.9~\cite{chang2023molecular} & 14.0~\cite{tholke2022torchmd} & 7924.8~\cite{tran2022open} & 703.9,~\cite{liao2022equiformer} 705.0~\cite{shoghi2024distribution} \\
  EquiformerV2~\cite{liao_equiformerv2_2024} &  & 6.17~\cite{liao_equiformerv2_2024} & 659.8~\cite{liao_equiformerv2_2024} & 316.0~\cite{liao_equiformerv2_2024} \\
  SpinConv~\cite{shuaibi2021rotation} &  & 12.0~\cite{shuaibi2021rotation} & 1944.0~\cite{tran2022open} & 437.0,~\cite{liao_equiformerv2_2024} 673.8~\cite{shoghi2024distribution}\\
  GemNet-dT~\cite{gasteiger2021gemnet} &  &  & 1271.3~\cite{tran2022open} & 400.0~\cite{liao_equiformerv2_2024}  \\
  GemNet-OC~\cite{gasteiger2021gemnet} &  &  & 828.7~\cite{tran2022open} & 344.0,~\cite{shoghi2024distribution} 355.0~\cite{liao_equiformerv2_2024} \\
  NequIP~\cite{batzner20223} & 4.2,~\cite{batzner20223} 0.7*~\cite{batatia2022mace} &  &  & 736.0~\cite{shoghi2024distribution} \\
  Equiformer~\cite{liao2022equiformer} &  & 6.59~\cite{liao2022equiformer} &  & 603.0,~\cite{kim2024geotmi} 630.1~\cite{liao2022equiformer} \\
  Faenet~\cite{duval2023faenet} & & 6.79~\cite{duval2023faenet} & & 551.0~\cite{ramlaoui2024improving} \\
  SEGNN~\cite{yu2022graph} &  & 15.0~\cite{liao2022equiformer} &  & 679.0~\cite{shoghi2024distribution} \\
  SphereNet~\cite{coors2018spherenet} &  & 6.0~\cite{liao2022equiformer} &  & 637.8~\cite{shoghi2024distribution} \\
  CGCNN~\cite{xie2018crystal} &  &  &  & 851.0~\cite{shoghi2024distribution} \\ 
  Allegro~\cite{musaelian2023learning} & 0.8*~\cite{batatia2022mace} & 4.7~\cite{shoghi2023molecules} &  &  \\
  EAA~\cite{chang2023molecular} & 5.9~\cite{chang2023molecular} & 12.0~\cite{chang2023molecular} &  &  \\
  GM-sNN~\cite{zaverkin2020gaussian} & 7.1,~\cite{chang2023molecular} 21.2~\cite{kovacs2021linear} & 11.7~\cite{zaverkin2020gaussian} &  &  \\
  MACE~\cite{batatia2022mace} & 0.9*~\cite{batatia2022mace} & 4.1~\cite{shoghi2023molecules} &  &  \\
  PhysNet~\cite{unke2019physnet} & 5.3~\cite{zhang2022reann} & 8.2~\cite{tholke2022torchmd} &  &  \\
  TorchMD-NET~\cite{tholke2022torchmd} & 3.6~\cite{tholke2022torchmd} & 6.2~\cite{tholke2022torchmd} &  &  \\
  ANI~\cite{smith2017ani} & 21.7,~\cite{kovacs2021linear} 8.8*~\cite{batatia2022mace} &  &  &  \\
  ACE~\cite{kovacs2021linear} & 2.0,~\cite{kovacs2021linear} 2.2*~\cite{batatia2022mace} &  &  &  \\
  BOTNet~\cite{batatia2022design} & 0.7*~\cite{batatia2022mace} &  &  &  \\
  EANN~\cite{zhang2019embedded} & 6.4,~\cite{zhang2022reann} 6.5~\cite{chang2023molecular} &  &  &  \\
  FCHL~\cite{faber2018alchemical} & 4.6,~\cite{zhang2022reann} 5.2,~\cite{kovacs2021linear} 2.0*~\cite{batatia2022mace} &  &  &  \\
  GAP~\cite{bartok2010gaussian} & 16.1,~\cite{kovacs2021linear} 6.0*~\cite{batatia2022mace} &  &  &  \\
  NewtonNet~\cite{haghighatlari2022newtonnet} & 5.0~\cite{batzner20223} &  &  &  \\
  REANN~\cite{zhang2022reann} & 4.4~\cite{zhang2022reann} &  &  &  \\
  sGDML~\cite{chmiela2018towards} & 5.0,~\cite{zhang2022reann} 6.9~\cite{kovacs2021linear} &  &  &  \\
  So3krates~\cite{frank2022so3krates} & 4.3~\cite{frank2022so3krates} & & & \\
  Cormorant~\cite{anderson2019cormorant} &  & 22.0~\cite{tholke2022torchmd} &  &  \\
  EGNN~\cite{satorras2021n} &  & 11.0~\cite{tholke2022torchmd} &  &  \\
  EQGAT~\cite{le2022equivariant} &  & 25.0~\cite{liao2022equiformer} &  &  \\
  LieConv~\cite{finzi2020generalizing} &  & 19.0~\cite{shuaibi2021rotation} &  &  \\
  NMP~\cite{gilmer2017neural} &  & 20.0~\cite{shuaibi2021rotation} &  &  \\
  \end{tabularx}
  
  \caption{Mean absolute error (MAE/meV) for the prediction of energies (and geometry optimisations in case of OC20 IS2RE) using popular {\ML} methods. {\ML} methods trained and tested on the MD17/rMD17* database (1,000 training data points; rMD17 has improved convergence setting for the underlying training data), the QM9 database (110,000 training data points), the OC20 database (460,328 training data points), and the OC22 database (8,225,293 training data points).}
\label{tab:MLIP_accuracy}
\end{table*}

\clearpage

\end{document}